\documentclass[10pt,journal,twocolumn,twoside]{IEEEtran}
\usepackage{setspace}
\usepackage{cite}
\usepackage{color}
\usepackage{amsmath}
\usepackage{amsfonts}
 \usepackage{amssymb}
\usepackage{graphicx}
\usepackage{algorithm}
\usepackage{algorithmic}
\usepackage{cleveref}
\usepackage{epsfig}
 \usepackage{epstopdf}
\usepackage{url}
 \usepackage[noblocks]{authblk}
 \usepackage{makecell}
 \usepackage{verbatim}
\begin{document}
%\begin{spacing}{2}

\title{Energy-Efficient Port Selection and Beamforming Design for Integrated Data and Energy Transfer Assisted by Fluid Antennas}
%\author{Author 1, Author 2, Author 3, Author 4}

\author{Long Zhang,~\IEEEmembership{Graduate Student Member ,~IEEE}, Yizhe Zhao,~\IEEEmembership{Member,~IEEE}, Halvin Yang,~\IEEEmembership{ Member,~IEEE}, Guangming Liang,~\IEEEmembership{Graduate Student Member ,~IEEE}, Jie Hu,~\IEEEmembership{Senior Member,~IEEE}
\thanks{This work was supported in part by the Natural Science Foundation of China (NSFC) under Grant No.62571091, No.62201123 and No.62431002 in part by the Young Elite Scientists Sponsorship Program by CAST under Grant 2023QNRC001, in part by the Natural Science Foundation of Sichuan (NSFSC) under Grant 2024NSFSC1417, and in part by Zhejiang Province Major Research and Development Plan (No.2024C01062).}
\thanks{ Long Zhang and Jie Hu are with the School of Information and Communication Engineering, University of Electronic Science and Technology of China, Chengdu 611731, China (e-mail: l.zhang@std.uestc.edu.cn;  hujie@uestc.edu.cn). }
\thanks{Yizhe Zhao is with the School of Information and Communication Engineering, University of Electronic Science and Technology of China, Chengdu 611731, China and he is also affiliated with the Yangtze Delta Region Institute (Huzhou), University of Electronic Science and Technology of China, Huzhou 313001, China (e-mail:yzzhao@uestc.edu.cn).}
\thanks{Halvin Yang is with the Wolfson School of Mechanical, Electrical and
	Manufacturing Engineering, Loughborough University, Loughborough, LE113TU, United Kingdom (e-mail: h.yang6@lboro.ac.uk).}
\thanks{Guangming Liang is with the School of Computing Science, University of Glasgow, G12 8RZ Glasgow, U.K. ( e-mail: g.liang.1@research.gla.ac.uk).}

}

%Email: l.zhang@std.uestc.edu.cn, uceehhy@ucl.ac.uk, yzzhao@uestc.edu.cn, hujie@uestc.edu.cn
\maketitle

\vspace{-2cm}
%\vspace{-1.6cm}

\begin{abstract}
Integrated data and energy transfer (IDET) is considered as a key enabler of 6G, as it can provide both wireless energy transfer (WET) and wireless data transfer (WDT) services towards low power devices. Thanks to the extra degree of freedom provided by fluid antenna (FA), incorporating FA into IDET systems presents a promising approach to enhance energy efficiency performance. This paper investigates a FA assisted IDET system, where the transmitter is equipped with multiple FAs and transmits wireless signals to the data receiver (DR) and the energy receiver (ER), both of which are equipped with a single traditional antenna. The switching delay and energy consumption induced by port selection are taken into account in IDET system for the first time. We aim to obtain the optimal beamforming vector and the port selection strategy at the transmitter, in order to maximize the short-term and long-term WET efficiency, respectively. The instant sub-optimal solution is obtained by alternatively optimizing the beamforming vector and port selection in each transmission frame, while a novel constrained soft actor critic (C-SAC) algorithm is proposed to find the feasible policy of port selection from the long-term perspective. Simulation results demonstrate that our scheme is able to achieve greater gain in terms of both the short-term and long-term WET efficiency  compared to other benchmarks, while not degrading WDT performance.
\end{abstract}


% \fi
\IEEEpeerreviewmaketitle

\section{Introduction}
\subsection{Background}
 Device proliferation and dramatic traffic explosion will occur in the Internet-of-Things (IoT) era, incurring the scarcity of spectrum resources and more power consumption on wireless devices \cite{9390169}. Furthermore, most of IoT devices are energy-constrained and rely on finite battery capacity to meet their sensing, processing, and transmission needs. Limited energy resources necessitate frequent charging or battery replacement, leading to increased operational costs. Integrated data and energy transfer (IDET) is an emerging paradigm to alleviate the problem of high energy consumption for the dual purpose of wireless data transfer (WDT) and wireless energy transfer (WET) by exploiting the controllable radio frequency (RF) signals \cite{10534278}. However, the low efficiency caused by the transmission attenuation of RF signals is a long standing issue in IDET systems.

In recent years, fluid antennas (FAs) have been envisioned as a promising technology for enhancing both the spectral and energy efficiency of wireless communication networks. Fluid antennas refer to any software-controllable fluidic, conductive or dielectric radiating structure that can change their shape or position to reconfigure the operating frequency, radiation pattern and other characteristics \cite{9264694}, which can provide extra degrees of freedom than traditional multiple-input multiple-output (MIMO). Unlike traditional fixed antenna assisted wireless networks, the fluid antenna system (FAS) can enhance the target signal strength and alleviate the impact of interference through the position (also known as port) adjustment of transmit and/or receive antenna and flexible beamforming. With the deployment of FA in IDET system, both the WDT performance and WET performance can be further improved.

Physical design of FA mainly focuses on liquid-based antennas, reconfigurable pixel-based antennas and stepper motor based antennas \cite{10146274}. The liquid based FA consists of liquid radiating elements (e.g., mercury, ga, galinstan, etc.) enclosed within a dielectric holder and controlled by a dedicated microelectromechanical system (MEMS) \cite{9992289}. The reconfigurable pixel based FA is composed of an array of metasurface elements or reconfigurable sub-wavelength structures that can dynamically modify their electromagnetic properties \cite{9743796}\cite{10201473} and the stepper motor based FA is known as the movable antenna (MA)\cite{10286328}. Compared to the high hardware costs of pixel-based antennas, which stem from a large number of fixed antennas, liquid-based FAs and MAs are more cost-effective as they require fewer antenna elements and RF chains. However, liquid-based FAs and MAs incur delay and energy consumption during the movement, which potentially limiting their use in some communication scenarios. Therefore, evaluating the impact of FA movement is crucial for achieving higher energy efficiency and spectral efficiency of IDET systems.

\subsection{Related works}
In order to establish an energy-efficient interconnected network, IDET  leverages RF signals to supply energy for low-power wireless devices. Plenty of works have been dedicated to the transceiver design of IDET system, aiming to improve the performance trade-off between WDT and WET.  For instance, Kwon \textit{et al.}  \cite{9454368} developed a transceiver design for a mmWave IDET system, including the transmit power allocation, beamforming and receive power splitting. Zhang \textit{et al.} \cite{10500404} considered a practical IDET  scenario, where the data receivers (DRs) and energy receivers (ERs) were located in the far- and near-field regions of the extremely large-scale array base station (BS), respectively.  The weight sum-power harvested at ERs  was maximized by jointly optimizing the beamforming vectors and transmit power under the constraints of maximum sum-rate. With the assistance of reconfigurable intelligent surface (RIS) \cite{9133435,10115291,11139123}, the performance of IDET system can be further improved.  Wu \textit{et al.} \cite{9133435} studied a sets of RIS assisted IDET system, where the beamforming vectors and reflect phase shifts were jointly optimized to minimize the transmit power, subject to the quality-of-service (QoS) constraints for all users. Yaswanth \textit{et al.}  \cite{10115291} jointly optimized the beamforming matrices at both base station and RIS to minimize the transmit power while satisfying the minimum DR and ER constraints, thereby contributing to an energy efficient transmission scheme. More recently, with the emergence of integrated sensing and communication (ISAC) \cite{10382465,zhous2025,pengs2025}, efforts have been made to integrate ISAC into IDET systems, offering further opportunities for energy-efficient and multifunctional network designs.

In recent years, FAS has attracted significant attention due to their exceptional flexibility and reconfigurability.  Research on FAS is still in its infancy and many studies focus on the performance analysis and beamforming design of FAS. Specifically, the concept of FAS was first proposed in \cite{9264694}, in which the performance of outage probability was proved to be superior to traditional maximum ratio combining (MRC) scheme. In addition, Wong \textit{et al.} proposed fast fluid antenna multiple access (f-FAMA) \cite{9650760} and slow fluid antenna multiple access (s-FAMA) \cite{10066316} to study FA assisted multiuser communications. Furthermore, a specific version of FAMA was proposed in \cite{10318083}, where a dense and fixed massive antenna array was deployed at each user. While some research focuses on the performance analysis of FAS, others develop optimization algorithms to enhance the practical performance of FAS. For instance, Efrem \textit{et al.} \cite{10677535} maximized the channel capacity in fluid MIMO system, where the port selection of both transmitter and receiver were optimized. In \cite{10473750}, the transmitter beamforming vectors and antenna position were jointly optimized to maximize the sum-rate in multiuser communications system, which demonstrated the superior performance of FA.

Owning to the superiority of FA, a few works \cite{10622204,lin2024,10615443,10840319,10506795} have integrated FA into IDET system, achieving better WET performance. Specifically, the outage probabilities of WDT and WET were analyzed in FA assisted IDET system by conceiving the time switching (TS) approach \cite{10622204} and power splitting (PS) approach \cite{lin2024}. Lai \textit{et al.} \cite{10615443} studied a FAS assisted wireless power communication system, where the analytical and asymptotic outage probability were derived. In addition, Zhou \textit{et al.} \cite{10840319} studied the multiple input and single output (MISO) FA assisted IDET system, where all antennas' position and the transmit beamforming vectors were optimized to maximize the communication rate of DR. In \cite{10506795}, the weight energy harvesting  power of ERs was maximized by jointly optimize the receivers' port selection and transmitter's beamforming vectors, demonstrating that a larger number of ports and a larger antenna size lead to improved IDET performance.
\subsection{Motivation and Contributions}

	Although previous works, including our own work \cite{10506795} have shown that deploying FAs in IDET systems can achieve outstanding performance than traditional antennas, they have not considered the impact of switching delay and energy consumption associated with antenna movement, which may decrease the throughput and energy efficiency of IDET system. In IDET system, the WDT performance and WET performance can be improved by moving FAs into strategic positions, where the signal strength is maximized, or interference is minimized, or a trade-off between both factors is achieved. However, the movement of FAs introduces both delay and energy consumption, which reduces the effective transmission time within each frame, thereby lowering the achievable throughput of DR and harvested energy of ER. Therefore, it is crucial to balance the channel gain brought from the FAs and the energy efficiency loss caused by their movement. Motivated by this, we investigate the impact of delay and energy consumption due to FA movement on the IDET performance from both short-term and long-term perspectives. To improve the energy efficiency of FA-assisted IDET system, we optimize beamforming design and port selection to maximize the WET efficiency at the ER. However, this energy-oriented design may misalign with the optimal direction for the DR, potentially compromising WDT performance. To address this trade-off, we incorporate a minimum throughput constraint at the DR in both short-term and long-term optimization, ensuring reliable communication while enhancing WET efficiency. The contributions are summarized as follows:
\begin{itemize}
	\item We study a FA assisted IDET system with the novel consideration of switching delay and energy consumption of the FA, where the transmitter is equipped with multiple FAs and the DR as well as ER are each equipped with a single traditional antenna. The port switching delay and energy consumption model of FA are considered in this paper for the first time.
	\item The WET efficiency is maximized by optimizing the beamforming vector and port selection from both short- and long-term perspective, while ensuring the harvested energy constraint and achievable throughput constraint. In short-term optimization, we transform the original fraction problem into linear problem by using Dinkebach's  algorithm, while the port selection and beamforming vector are then optimized  alternatively by using feasible point pursuit-successive convex approximation (FPP-SCA) and semidefinite relaxation (SDR).
	\item In long-term optimization, we propose a novel constrained soft actor critic (C-SAC) algorithm to tackle the transformed zero-sum Markov Bandit game, in order to find the  feasible policy of port selection. The optimal long-term WET efficiency is   then search by adopting bisection search method.
	\item Simulation results demonstrate that the short-term WET efficiency of our proposed algorithm outperforms the benchmark that does not consider the impact of delay and energy consumption. Furthermore,  with the assistance of deep reinforcement learning (DRL) agent, we can achieve a better long-term WET efficiency compared to short-term optimization.
\end{itemize}

The rest of this paper is organized as follows. The model of FA assisted IDET system is introduced in Section \ref{section:architecture}. The short-term optimization and long-term optimization for WET efficiency are studied in Section \ref{short} and Section \ref{long}, respectively. Simulation results are provided in Section \ref{section:sim} and finally, our paper is concluded in Section \ref{section:conclusion}.

\textit{Notations:}
In this paper, boldface upper letters, boldface lower letters, and lower letters denote matrices, vectors, and scalars, respectively. $\mathbb{C}^{x\times y}$ and $\mathbb{R}^{x\times y}$ represent the set of complex matrices and real matrices with the dimension of $x\times y$, respectively. $|x|$ represents the amplitude of a complex number $x$. $||\boldsymbol{x}||$ and $||\boldsymbol{x}||_1$ denote the 2-norm and 1-norm of a vector $\boldsymbol{x}$, respectively. The notation  $(\cdot)^T$ and $(\cdot)^H$ refer to the transpose and the conjugate transpose of a vector or matrix, while $\mathbf{E}$ represents the expectation operator.

\section{System Model}\label{section:architecture}
\begin{figure}
	\centering
	\includegraphics[width=0.6\linewidth]{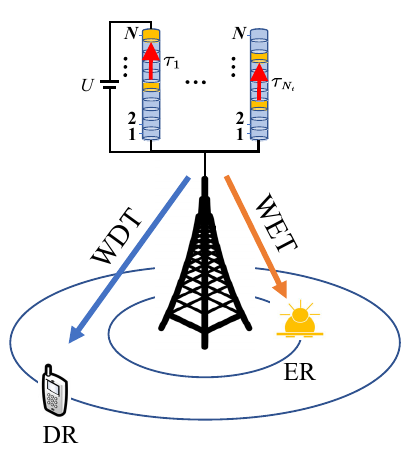}
	\caption{System model.}
	\label{fig:system_model}
\end{figure}

As shown in Fig. \ref{fig:system_model}, we consider a multiple FAs assisted IDET system, which is comprised of a transmitter, a DR and an ER. The transmitter is equipped with $N_t$ FAs, each of which has the same size and parameters without loss of generality. The DR and ER are both equipped with a single traditional antenna due to the limited hardware space. We assume that the transmit antennas are sufficiently spaced  (larger than $\lambda/2$ or equal to $\lambda/2$  ), such that their spatial correlation and mutual coupling can be considered negligible \cite{10677535}. Consequently, the channels of distinct fluid antennas are independent, consistent with the assumptions commonly used in classic MIMO systems. However, the channels among different ports in the same antenna are strongly spatially correlated due to their close physical proximity. The DR and ER are randomly distributed in two different directions of the transmitter, which allows the beamforming vector to be optimized with sufficient spatial flexibility. Moreover, considering that the ER typically requires a higher harvested energy to meet its operational threshold, it is generally deployed closer to the transmitter to compensate for the severe pathloss in wireless energy transfer. The set of FAs and FA's port are denoted by $\mathcal{N}_t=\left\{1,\cdots,N_t\right\}$ and $\mathcal{N}=\left\{1,\cdots,N\right\}$, respectively.

\subsection{Fluid antenna model}
In this paper, we consider the practical architecture of the liquid metal-based fluid antenna. The length of fluid antenna is $W\lambda$, where $\lambda$ is the wavelength of RF signal and $W$ is a scaling constant. The MEMS is adopted in each fluid antenna to drive the metal droplet to move. Specifically, due to the electrocapillary effect, the motion of metal droplet is achieved by applying a voltage gradient along the FA, resulting in a phenomenon known as continuous electrowetting \cite{10163953}. According to \cite{9992289}, the average velocity of the metal droplet motion is given by
\begin{align}
	u=\frac{q}{6\mu}\frac{D}{L}\Delta\phi,
\end{align}
where $q$  is the initial charge in the electrical double layer for EGaIn, $\mu$ denotes the viscosity of EGaIn, $D$ and $L$ are the thickness and length of metal droplet, respectively. $\Delta\phi$ represents the voltage differential resulting from the flow of current through the narrow electrolyte layer between EGaIn and the wall, which is much smaller than the voltage $U$ applied  between two ends of the FA \cite{846697}. Therefore, the moving delay from $i$-th port to $j$-th port can be expressed as
\begin{align}
	\tau=\frac{W\lambda}{u}\frac{|i-j|}{N-1}.
\end{align}

\subsection{Wireless channel model}
The block fading channel model is assumed between the transmitter and receivers, implying that the channel remains constant during a coherence time $T$ but varies from one frame to another. Since the ports are located much closer in the same FA, strong correlation exists between the channels of different ports. Following Jake's two dimensional correlation model \cite{10103838}, the correlation between any two ports can be represented by a matrix $\boldsymbol{J}$, which is given by
\begin{align}
	\boldsymbol{J}=\left[\begin{matrix}
		J_{1,1} &J_{1,2} &\cdots &J_{1,N}\\
		J_{2,1}&J_{2,2} &\cdots &J_{2,N}\\
		\vdots &\vdots &\ddots&\vdots\\
		J_{N,1}&J_{N,2}&\cdots&J_{N,N}
	\end{matrix}\right],
\end{align}
 where $J_{i,j}=J_0(\frac{2\pi(i-j)}{N-1}W)$ and $J_0(\cdot)$ is the zeroth-order Bessel function of the first kind. The channels between $i$-th FA and DR and ER in the $t$-th block are denoted as $\boldsymbol{h}_i(t)$ and $\boldsymbol{g}_i(t)$, which can be modeled as
  \begin{align}\label{spa}
   \boldsymbol{h}_i(t)=\sqrt{\frac{1}{PL_{h,i}}}\boldsymbol{U}\sqrt{\boldsymbol{\Lambda}}\boldsymbol{z}_h(t),\notag\\
   \boldsymbol{g}_i(t)=\sqrt{\frac{1}{PL_{g,i}}}\boldsymbol{U}\sqrt{\boldsymbol{\Lambda}}\boldsymbol{z}_g(t),
  \end{align}
where $PL_{h,i}$ and $PL_{g,i}$ are the path-loss from $i$-th FA to DR and ER, respectively. $\boldsymbol{U}$ and $\boldsymbol{\Lambda}$ are a unitary matrix composed of the eigenvectors of $\boldsymbol{J}$ and a diagonal matrix consisting of the eigenvalues of $\boldsymbol{J}$, which satisfy $\boldsymbol{J}=\boldsymbol{U}\boldsymbol{\Lambda}\boldsymbol{U}^{T}$. In addition,   $\boldsymbol{z}_h(t)\in\mathbb{C}^{N\times 1}$ and $\boldsymbol{z}_g(t)\in\mathbb{C}^{N\times 1}$ are complex Gaussian random vector with each element having zero mean and variance of 1.

On the other hand, the channels between the transmitter and receivers are time-varying and exhibit temporal correlation across transmission frames. A widely accepted method to model channel evolution is to adopt a first-order Gaussian-Markov process with a one-step correlation coefficient derived from Jakes' model \cite{10478288,10078846}. Specifically, the first-order  Gaussian-Markov channel model can be expressed as
\begin{align}\label{tem}
	\boldsymbol{h}_i(t)=\rho\boldsymbol{h}_i(t-1)+\sqrt{\frac{1-\rho^2}{PL_{h,i}}}\boldsymbol{\delta}(t),\notag\\
	\boldsymbol{g}_i(t)=\rho\boldsymbol{g}_i(t-1)+\sqrt{\frac{1-\rho^2}{PL_{g,i}}}\boldsymbol{\delta}(t)
\end{align}
where $\rho\in [0,1]$ is the temporal correlation coefficient and $\boldsymbol{\delta}\sim\mathcal{CN}(0,\boldsymbol{I}_N)$ is the random difference with unit-variance complex Gaussian i.i.d. in time.
\subsection{Signal model}
 Before the transmitter delivering wireless signal to the DR and ER, an appropriate antenna port should be selected on each FA. Assuming that the initial port of $i$-th FA is $k_{i,0}(t)$ in the $t$-th block, the port switching delay on the $i$-th antenna from the initial port to the selected port $k_{i}(t)$ is expressed as
\begin{align}
	{\tau}_i(t)=\frac{W\lambda}{u}\frac{|k_i(t)-k_{i,0}(t)|}{N-1}.
\end{align}
Therefore, the time consumption in port selection phase can be expressed by
\begin{align}
	\tau_s(t)=\max\{\tau_1(t),\cdots,\tau_{N_t}(t)\}.
\end{align}
We assume that the transmitter only transmits signal for the DR, while the ER can harvest energy from the transmitted signal. Denoted the beamforming vector and the selected channel as $\boldsymbol{w}(t)=[w_1(t),w_2(t),\cdots,w_{N_t}(t)]^T$ and $\boldsymbol{h}_s(t)=[{h}_1(t,k_1(t)),\cdots,{h}_{N_t}(t,k_{N_t}(t))]^T$ as well as $\boldsymbol{g}_s(t)=[{g}_1(t,k_1(t)),\cdots,{g}_{N_t}(t,k_{N_t}(t))]^T$, where ${h}_i(t,k_i(t))$ and ${g}_i(t,k_i(t))$ represent the channels between the $k_{i}(t)$-th selected port on the $i$-th FA and DR as well as ER, the received signal by DR is given by
\begin{align}
	\boldsymbol{y}_D(t)=\boldsymbol{h}_s(t)^H\boldsymbol{w}(t)s(t)+n(t),
\end{align}
where $n(t)$ is the additive white Gaussian noise (AWGN) at DR having zero mean and the variance of $\sigma^2$ and $s(t)$ is the Gaussian distributed information bearing signal having unit power. Therefore, the achievable throughput  at DR in each frame is expressed as
\begin{align}
	R(t)=\frac{T-\tau_s(t)}{T}\log(1+\frac{|\boldsymbol{h}_s(t)^H\boldsymbol{w}(t)|^2
	}{\sigma^2}).
\end{align}

On the other hand, by ignoring the noise power, the harvested energy at ER is given by
\begin{align}
	E_h(t)=\eta(T-\tau_s(t))|\boldsymbol{g}_s(t)^H\boldsymbol{w}(t)|^2,
\end{align}
where $\eta\in(0,1]$ represents the energy conversion efficiency\footnote{ In general, considering a nonlinear energy harvest model is more practical, but the received power at ER is much small due to the signal attenuation, enabling the EH circuit works in linear region of the nonlinear model \cite{10185576}. Thus the linear EH model can be used to approximate the nonlinear model.}.  Assuming that the power actuating the liquid metal moving is a constant as $P_u$, the total energy consumption in coherence time is expressed as
\begin{align}
	E_c(t)&=(T-\tau_s(t))||\boldsymbol{w}(t)||^2+\sum_{i=1}^{N_t}\tau_i(t)P_u+ T P_c,
\end{align}
where $P_c$ is the circuit power required to maintain the normal operation of the system. Therefore, the WET efficiency can be expressed as
\begin{align}
	&\eta_{EE}(t)=\frac{E_h(t)}{E_c(t)}\notag \\
	=&\frac{\eta(T-\tau_s(t))|\boldsymbol{g}_s(t)^H\boldsymbol{w}(t)|^2}{(T-\tau_s(t))||\boldsymbol{w}(t)||^2+\sum_{i=1}^{N_t}\tau_i(t)P_u+ T P_c}.
\end{align}

In this paper, we aim to investigate both short-term and long-term WET efficiency maximization problem for the FA assisted IDET system,  subject to the throughput constraint of DR. Specifically:
\begin{itemize}
	\item The short-term optimization focuses on instant WET efficiency improvements for each transmission frame with relatively low complexity, which is more suitable for fast-changing environment. However, its myopic optimization horizon may lead to suboptimal performance accumulation over extended operational periods.
	\item Conversely, the long-term optimization framework adopts a holistic perspective by considering temporal correlations of channels and state evolution across multiple transmission frames, which can achieve more stable long-term WET efficiency. Nevertheless, the long-term optimization based DRL algorithm requires prior experience and the training is time-consuming.
\end{itemize}
 Note that when port switching delay and energy consumption are explicitly incorporated into the WET efficiency optimization, the system must balance the channel gain brought by port reconfigurability against the efficiency loss caused by antenna movement. This trade-off substantially increases the problem complexity and highlights the need for dedicated cost-aware optimization frameworks. 

\section{Short-term Energy-Efficient Port Selection and Beamforming Design}\label{short}
\subsection{Problem Formulation}
In this section, we optimize the beamforming vector and port selection to maximize the WET efficiency in each transmission frame by only exploiting the current channel state information (CSI). For the sake of simplicity, we omit the time index $t$ in the expression. Therefore, the optimization problem can be formulated as
 \begin{align}
	\text{(P1)}:\max_{\boldsymbol{w},\{k_i\}} \quad & \eta_{EE}, \label{obj0}\\
	s.t.  \ \ \quad &E_h\ge E_{th}, \tag{\ref{obj0}{a}}\label{obj0a}\\
	&R\ge R_{th}, \tag{\ref{obj0}{b}}\label{obj0b}\\
	&||\boldsymbol{w}||^2\leq P_{max}, \tag{\ref{obj0}{c}}\label{obj0c}\\
	&\tau_s\leq T, \tag{\ref{obj0}{d}}\label{obj0d}
\end{align}
where \eqref{obj0a} and \eqref{obj0b} guarantee the minimum achievable throughput constraint and minimum harvested energy constraint, respectively. \eqref{obj0c} imposes the maximum transmission power $P_{max}$ for the transmitter. In order to solve this problem, we introduce port activation indicators $\{\boldsymbol{a}_i\in \mathbb{R}^{N\times 1}\}$ to denote the port selection of FAs, where  $\boldsymbol{a}_i$ is the $i$-th FA's port activation indicator and only one element is 1 and the rest is 0.   Defined $\hat{\boldsymbol{a}}=[\boldsymbol{a}_1^T,\cdots,\boldsymbol{a}_{N_t}^T]^T$ as well as an operation of converting vectors into matrix as
\begin{align}
	\boldsymbol{H}=	\mathcal{M}\left\{\boldsymbol{h}_i,i\in\mathcal{N}_t\right\}&=\left[\begin{matrix}
		\boldsymbol{h}_1 &0 &\cdots &0\\
		0&\boldsymbol{h}_2 &\cdots &0\\
		\vdots &\vdots &\ddots&\vdots\\
		0&0&\cdots&\boldsymbol{h}_{N_t}
	\end{matrix}\right],
\end{align} the selected channels between transmitter and the DR as well as the ER can be expressed as $\boldsymbol{h}_s^H=[\boldsymbol{a}_1^T\boldsymbol{h}_1,\cdots,\boldsymbol{a}_{N_t}^T\boldsymbol{h}_{N_t}]=\hat{\boldsymbol{a}}^T\boldsymbol{H}$ and $\boldsymbol{g}_s^H=[\boldsymbol{a}_1^T\boldsymbol{g}_1,\cdots,\boldsymbol{a}_{N_t}^T\boldsymbol{g}_{N_t}]=\hat{\boldsymbol{a}}^T\boldsymbol{G}$, respectively. In addition, denoted the port switching delay vector of $i$-th FA from the initial port to other ports as
\begin{align}
\boldsymbol{d}_i=[\frac{W\lambda}{u}\frac{|1-k_{i,0}|}{N-1},\cdots,\frac{W\lambda}{u}\frac{|N-k_{i,0}|}{N-1}]^T,
\end{align}
the port switching delay at $i$-th FA can be given by $\tau_i=\boldsymbol{d}_i^T\boldsymbol{a}_i$. Therefore, (P1) can be reformulated as
\begin{align}
 	\text{(P2)}:\max_{\boldsymbol{w},\hat{\boldsymbol{a}},\tau_s} \quad & \frac{\eta(T-\tau_s)|\hat{\boldsymbol{a}}^T\boldsymbol{G}\boldsymbol{w}|^2}{(T-\tau_s)||\boldsymbol{w}||^2+||\hat{\boldsymbol{a}}^T\boldsymbol{D}||_1P_u+ T P_c}, \label{obj1}\\
 	s.t.  \ \ \quad &\eta(T-\tau_s)|\hat{\boldsymbol{a}}^T\boldsymbol{G}\boldsymbol{w}|^2\ge E_{th}, \tag{\ref{obj1}{a}}\label{obj1a}\\
 	&\frac{T-\tau_s}{T}\log(1+\frac{|\hat{\boldsymbol{a}}^T\boldsymbol{H}\boldsymbol{w}|^2
 	}{\sigma^2})\ge R_{th}, \tag{\ref{obj1}{b}}\label{obj1b}\\
 	&||\boldsymbol{w}||^2\leq P_{max}, \tag{\ref{obj1}{c}}\label{obj1c}\\
 	&\tau_s\leq T, \tag{\ref{obj1}{d}}\label{obj1d}\\
 	&\hat{\boldsymbol{a}}^T\boldsymbol{D}\leq\tau_s\boldsymbol{1}^{N_t},\tag{\ref{obj1}{e}}\label{obj1e}\\
 	&\hat{\boldsymbol{a}}_{k}=0\ \mathrm{or}\ 1,\notag\\
 	&\sum_{k=1}^{N} \boldsymbol{a}_{i,k}=1 , i=1,\cdots,N_t,\tag{\ref{obj1}{f}}\label{obj1f}
\end{align}
 where $\boldsymbol{G}=\mathcal{M}\left\{\boldsymbol{g}_i,i\in\mathcal{N}_t\right\}$ and $\boldsymbol{D}=\mathcal{M}\left\{\boldsymbol{d}_i,i\in\mathcal{N}_t\right\}$. Note that the objective function is a complicated fraction and the beamforming vector and the port activation indicators are intricately coupled, thus (P2) is non-convex optimization problem and is hard to solve directly. Referring to previous work \cite{8314727}, Dinkelbach's algorithm is widely applied to transform fractional objective function into linear function by  introducing an auxiliary variable $q$, making the optimization problem more tractable. Note that when the single-ratio problem is a concave-convex fractional programming, the Dinkelbach's algorithm can converge to the global optimal solutions. Therefore,  we aim to reformulate the objective function into a concave-minus-convex form in the subsequent transformation. By relaxing the binary constraints and denoting $\boldsymbol{C}=\mathcal{M}\left\{\boldsymbol{1}_i^{N},i\in\mathcal{N}_t\right\}$, (P2) can be reformulated as
 \begin{align}
\text{(P3)}:\max_{\boldsymbol{w},\hat{\boldsymbol{a}},\tau_s} \quad & E_h-q E_c, \label{obj2}\\
s.t. \ \ \quad &E_h\ge E_{th}, \tag{\ref{obj2}{a}}\label{ob21a}\\
	&R\ge R_{th}, \tag{\ref{obj2}{b}}\label{obj2b}\\
	&||\boldsymbol{w}||^2\leq P_{max}, \tag{\ref{obj2}{c}}\label{obj2c}\\
&\hat{\boldsymbol{a}}^T\boldsymbol{D}\leq\tau_s\boldsymbol{1}^{N_t},\tag{\ref{obj2}{d}}\label{obj2d}\\
&\tau_s\leq T, \tag{\ref{obj2}{e}}\label{obj2e}\\
&\hat{\boldsymbol{a}}^T\boldsymbol{C}=\boldsymbol{1}^{N_t},\tag{\ref{obj2}{f}}\label{obj2f}\\
&\hat{\boldsymbol{a}}\in[0,1]^{NN_t}\tag{\ref{obj2}{g}}.\label{obj2g}
\end{align}
In (P3), $q$ is iteratively updated by
\begin{align}
	q^n=\frac{E_h^n}{E_c^n}
\end{align}
where $n$ is the iteration index. For a fixed parameter $q$, (P3) is still a non-convex problem, thus we develop a AO algorithm to tackle it by solving $\hat{\boldsymbol{a}} $, $\tau_s$ and $\boldsymbol{w}$ iteratively.
 \raggedbottom
\subsection{Port selection optimization}
We first optimize the port activation indicators  $\left\lbrace \boldsymbol{a}_i\right\rbrace$ and port switching delay $\left\{\tau_i\right\}$ by fixing the beamforming vector $\boldsymbol{w}$. The objective function of (P3) can be reformulated as
\begin{align}
	&\eta|\hat{\boldsymbol{a}}^T\boldsymbol{G}\boldsymbol{w}|^2-q\left( ||\boldsymbol{w}||^2+ \frac{||\hat{\boldsymbol{a}}^T\boldsymbol{D}||_1P_u+ T P_c}{(T-\tau_s)}\right) \\ \notag
	&=\eta|\hat{\boldsymbol{a}}^T\boldsymbol{G}\boldsymbol{w}|^2-q\left( ||\boldsymbol{w}||^2+ \frac{||\hat{\boldsymbol{a}}^T\sqrt{\boldsymbol{D}}||_2^2P_u+ T P_c}{(T-\tau_s)}\right)
\end{align}
where $\sqrt{\boldsymbol{D}}$ represents the element-wise square root operation applied to matrix $\boldsymbol{D}$. Therefore, when the beamforming vector $\boldsymbol{w}$ is fixed, (P3) can reformulated as
 \begin{align}
	\text{(P4)}:\max_{\hat{\boldsymbol{a}},\tau_s} \quad & \eta|\hat{\boldsymbol{a}}^T\boldsymbol{G}\boldsymbol{w}|^2-q\left( \frac{||\hat{\boldsymbol{a}}^T\sqrt{\boldsymbol{D}}||_2^2P_u+ T P_c}{(T-\tau_s)}\right), \label{obj3}\\
	s.t.  \ \ \quad &\eta|\hat{\boldsymbol{a}}^T\boldsymbol{G}\boldsymbol{w}|^2\ge \frac{E_{th}}{T-\tau_s}, \tag{\ref{obj3}{a}}\label{obj3a}\\
	&|\hat{\boldsymbol{a}}^T\boldsymbol{H}\boldsymbol{w}|^2\ge \sigma^2\left( 2^{\left(\frac{TR_{th}}{T-\tau_s}\right)}-1\right)  , \tag{\ref{obj3}{b}}\label{obj3b}\\
	&\tau_s\leq T, \tag{\ref{obj3}{c}}\label{obj3c}\\
	&\hat{\boldsymbol{a}}^T\boldsymbol{D}\leq\tau_s\boldsymbol{1}^{N_t},\tag{\ref{obj3}{d}}\label{obj3d}\\
	&\hat{\boldsymbol{a}}^T\boldsymbol{C}=\boldsymbol{1}^{N_t},\tag{\ref{obj3}{e}}\label{obj3e}\\
	&\hat{\boldsymbol{a}}\in[0,1]^{NN_t}.\tag{\ref{obj3}{f}}\label{obj3f}
\end{align}
 (P4) contains non-convex terms $|\hat{\boldsymbol{a}}^T\boldsymbol{G}\boldsymbol{w}|^2$ and $|\hat{\boldsymbol{a}}^T\boldsymbol{H}\boldsymbol{w}|^2$ in both the objective function \eqref{obj3} and constraints \eqref{obj3a}-\eqref{obj3b}, while maintaining convexity in all other terms with respect to optimization variables $\hat{\boldsymbol{a}}$ and $\tau_s$. Although semidefinite relaxation (SDR) approach can be used to linearize the non-convex components, its application would induce significant dimensionality expansion of the optimization variables, rendering it computationally prohibitive for FA-assisted IDET system with large port number $N$. The FPP-SCA algorithm \cite{6954488} is proposed for approximately solving general quadratic constraint quadratic problems (QCQPs), while a feasible point is guaranteed by adding slack variables. For any $\hat{\boldsymbol{a}}$,$\hat{\boldsymbol{x}}$, we have $(\hat{\boldsymbol{a}}-\hat{\boldsymbol{x}})^T\boldsymbol{S}(\hat{\boldsymbol{a}}-\hat{\boldsymbol{x}})\ge0$, where $\boldsymbol{S}=\eta\boldsymbol{G}\boldsymbol{w}\boldsymbol{w}^H\boldsymbol{G}^H$. Expanding the right-hand side of the inequality, we obtain
\begin{align}
	\hat{\boldsymbol{a}}^T\boldsymbol{S}\hat{\boldsymbol{a}}\ge2Re\left\lbrace\hat{\boldsymbol{a}}^T\boldsymbol{S}\hat{\boldsymbol{x}} \right\rbrace -	\hat{\boldsymbol{x}}^T\boldsymbol{S}\hat{\boldsymbol{x}}
\end{align}
Therefore, by applying FPP-SCA on (P4), it can be reformulated as
 \begin{align}
	\text{(P5)}: \max_{\hat{\boldsymbol{a}},\tau_s,s_1,s_2}  &  2Re\left\lbrace\hat{\boldsymbol{a}}^T\boldsymbol{S}\hat{\boldsymbol{x}} \right\rbrace -	\hat{\boldsymbol{x}}^T\boldsymbol{S}\hat{\boldsymbol{x}}-p(s_1+s_2)\notag  \\
	&-q\left( \frac{||\hat{\boldsymbol{a}}^T\sqrt{\boldsymbol{D}}||_2^2P_u+ T P_c}{(T-\tau_s)}\right),  \label{obj4} \\
	s.t. \ \ &2Re\left\lbrace\hat{\boldsymbol{a}}^T\boldsymbol{S}\hat{\boldsymbol{x}} \right\rbrace -	\hat{\boldsymbol{x}}^T\boldsymbol{S}\hat{\boldsymbol{x}} +s_1\ge \frac{E_{th}}{T-\tau_s}, \tag{\ref{obj4}{a}}\label{{obj4}a}\\
	&2Re\left\lbrace\hat{\boldsymbol{a}}^T\boldsymbol{F}\hat{\boldsymbol{x}} \right\rbrace -	\hat{\boldsymbol{x}}^T\boldsymbol{F}\hat{\boldsymbol{x}} +s_2\ge \notag\\ & \sigma^2\left( 2^{\left(\frac{TR_{th}}{T-\tau_s}\right)}-1\right)  , \tag{\ref{obj4}{b}}\label{{obj4}b}\\
	&\tau_s\leq T, \tag{\ref{obj4}{c}}\label{obj4c}\\
	&\hat{\boldsymbol{a}}^T\boldsymbol{D}\leq\tau_s\boldsymbol{1}^{N_t},\tag{\ref{obj4}{d}}\label{{obj4}d}\\
	&\hat{\boldsymbol{a}}^T\boldsymbol{C}=\boldsymbol{1}^{N_t},\tag{\ref{obj4}{e}}\label{{obj4}e}\\
	&\hat{\boldsymbol{a}}\in[0,1]^{NN_t},\tag{\ref{obj4}{f}}\label{{obj4}f}\\
	&s_1\ge0,s_2\ge0,\tag{\ref{obj4}{g}}\label{obj4g}
\end{align}
where $p$ is slack penalty, $s_1$ and $s_2$ are slack variables, $\boldsymbol{F}=\boldsymbol{H}\boldsymbol{w}\boldsymbol{w}^H\boldsymbol{H}^H$. Given an initial point $\hat{\boldsymbol{x}}$,  (P5) is a convex optimization problem and can be optimally solved with the assistance of convex optimization tools. Due to the application of first-order Taylor expansion, the solution obtained from (P5) is the sub-optimal of (P4), and the objective value of (P5) provides a lower bound for (P4).
\subsection{Transmit beamforming optimization}
In this subsection, we optimize the transmit beamforming vector by fixing the port activation indicators. Therefore, (P3) can be reformulated as
 \begin{align}
	\text{(P6)}:\max_{\boldsymbol{w}} \quad & \eta|\hat{\boldsymbol{a}}^T\boldsymbol{G}\boldsymbol{w}|^2-q||\boldsymbol{w}||^2, \label{obj5}\\
	s.t.  \ \quad &\eta|\hat{\boldsymbol{a}}^T\boldsymbol{G}\boldsymbol{w}|^2\ge \frac{E_{th}}{T-\tau_s}, \tag{\ref{obj5}{a}}\label{obj5a}\\
	&|\hat{\boldsymbol{a}}^T\boldsymbol{H}\boldsymbol{w}|^2\ge \sigma^2\left( 2^{\left(\frac{TR_{th}}{T-\tau_s}\right)}-1\right)  , \tag{\ref{obj5}{b}}\label{obj5b}\\
	&||\boldsymbol{w}||^2\leq P_{max}.\tag{\ref{obj5}{c}}\label{obj5c}
\end{align}
Let $\boldsymbol{W}=\boldsymbol{w}\boldsymbol{w}^H$ and then it follows $\text{rank}(\boldsymbol{W})\leq 1$. Dropping the rank constraint on  $\boldsymbol{W}$, the SDR of (P6) is given by
 \begin{align}
	\text{(P7)}:\max_{\boldsymbol{W}} \quad & \text{tr}(\boldsymbol{AW})-q\text{tr}(\boldsymbol{W}), \label{obj6}\\
	s.t.   \ \quad &\text{tr}(\boldsymbol{AW})\ge \frac{E_{th}}{T-\tau_s}, \tag{\ref{obj6}{a}}\label{obj6a}\\
	&\text{tr}(\boldsymbol{BW})\ge \sigma^2\left( 2^{\left(\frac{TR_{th}}{T-\tau_s}\right)}-1\right)  , \tag{\ref{obj6}{b}}\label{obj6b}\\
	&\text{tr}(\boldsymbol{W})\leq P_{max},\tag{\ref{obj6}{c}}\label{obj6c}\\
	&\boldsymbol{W}\ge 0.\tag{\ref{obj6}{d}}\label{obj6d}
\end{align}
where $\boldsymbol{A}=\eta\boldsymbol{G}^H\hat{\boldsymbol{a}}\hat{\boldsymbol{a}}^H\boldsymbol{G}$ and $\boldsymbol{B}=\boldsymbol{H}^H\hat{\boldsymbol{a}}\hat{\boldsymbol{a}}^H\boldsymbol{H}$. According to \cite{10506795}, the optimal solution $\boldsymbol{W}^*$ of (P7) satisfies $\text{rank}(\boldsymbol{W}^*)\leq 3$. Since $\boldsymbol{W}^*\neq \boldsymbol{0}$, thus we have $\text{rank}(\boldsymbol{W}^*)=1$, which indicates that the rank constraint relaxation is tight and (P7) has the same optimal solution as (P6). By performing eigenvalue decomposition over $\boldsymbol{W}^*$, the optimal beamforming vector $\boldsymbol{w}^*$ is obtained. Note that when $q$ is fixed, the objective value of (P3) is non-decreasing by solving (P5) and (P7) alternately. Thus the proposed AO algorithm can converge to the sub-optimal solutions. The overall proposed algorithm for solving (P3) is summarized in Algorithm  \ref{power}. However, the optimized solution $\hat{\boldsymbol{a}}^*$ does not inherently possess binary properties. In order to recover the original port activation indicators, we choose the index of largest value of $\boldsymbol{a}_i^*$ as the selected port.

\begin{algorithm}[t]
    \renewcommand{\algorithmicrequire}{\textbf{Initialize:}}
	\renewcommand{\algorithmicensure}{\textbf{\textsc{Repeat}:}}
	\caption{The Proposed Algorithm for solving (P2)}
    \label{power}
    \begin{algorithmic}[1]
        \REQUIRE  Initialize the WET efficiency $q^{(0)}=0$ and generate the port activation indicator $\hat{\boldsymbol{x}}^{(0)}$ and the beamforming vector $\boldsymbol{w}^{(0)}$ randomly. Set the maximal iteration number $M$ and the tolerance threshold $\epsilon$ ;
	    \ENSURE For a given $q^{(m-1)}$
        \STATE  \textbf{repeat}: For a given $\boldsymbol{w}^{(t-1)}$

        \STATE \quad \textbf{repeat}: For a given $\boldsymbol{x}^{(n-1)}$, solve problem (P5) to     obtain $\hat{\boldsymbol{a}}^{(n)}$ and $\tau_s^{(n)}$;
        \STATE \quad set $\boldsymbol{x}^{(n)}=\hat{\boldsymbol{a}}^{(n)}$ and $n=n+1$
        \STATE \quad\textbf{until} the objective value in (P5) converges;
        \STATE  Take $\hat{\boldsymbol{a}}^{(n)}$ into (P7) and solve (P7) to get the optimal solution $\boldsymbol{W}^{(t)}$. Perform EVD for $\boldsymbol{W}^{(t)}$ to obtain the beamforming vector $\boldsymbol{w}^{(t)}$. Set $t=t+1$.
        \STATE \textbf{until}  the objective value in (P3) converges;
        \STATE Update $q^{(m)}=\frac{E_h^{(m)}}{E_t^{(m)}}$ and set $m=m+1$.

        \STATE \textbf{\textsc{Until}} converges, i.e. $\frac{q^{(m)}-q^{(m-1)}}{q^{(m)}}\leq \epsilon$
        \STATE \textbf{{Output}} $\hat{\boldsymbol{a}}^*$ and $\boldsymbol{w}^*$
    \end{algorithmic}
\end{algorithm}

\subsection{Complexity analysis}
First, we denote the iteration number of the FPP-SCA algorithm for solving (P5) as $I_s$,  the iteration number of AO algorithm for solving (P3) as $I_a$ and the iteration number of Dinkelbach's algorithm as $I_d$. According to \cite{6954488} and \cite{9423652}, the computational complexity of FPP-SCA containing $x$ variables and $y$ constraints can be formulated as $\mathcal{O}(log(\frac{1}{\gamma})(x+y)^{3.5})$, while the computational complexity of SDP containing $x\times x $ positive semidefinite matrix and $y$ SDP constraints is $\mathcal{O}(\sqrt{x}log(\frac{1}{\gamma})(yx^3+x^2y^2+y^3))$, where $\gamma$ denotes the desired solution accuracy. Therefore, when the interior-point algorithm is applied to solve (P5) and (P7), the computational complexity of (P5) and (P7) can be expressed as $\mathcal{O}_1=\mathcal{O}(log(\frac{1}{\gamma})(NN_t+3)^{3.5})$ and $\mathcal{O}_2=\mathcal{O}(log(\frac{1}{\gamma})\sqrt{N_t}(3N_t^3+9N_t^2+27))$, respectively. The worst-case computational complexity  of the AO algorithm for solving (P2) can be obtained as $\mathcal{O}(I_dI_a(I_s\mathcal{O}_1+\mathcal{O}_2))$.

\section{Long-term Energy Efficient Port Selection and Beamforming Design}\label{long}

\subsection{Problem formulation and Transformation}
In the last section, we optimize the beamforming vector and port selection to maximize the instantaneous WET efficiency for every signal transmission frame. However, focusing solely on optimizing the port position in each transmission frame may result in a significant increase in delay and energy consumption in subsequent transmission frames, leading to a sharp decline in the long-term WET performance. Meanwhile, the optimal WET efficiency in the short term is not equal to the optimal WET efficiency in the long term. Therefore, in this section, we aim to maximize the average WET efficiency from the perspective of long-term optimization by using the DRL algorithm, while satisfying the long-term constraints of achievable throughput and harvested energy. The long-term optimization problem can be formulated as

 \begin{align}
	\text{(P8)}:\max_{\boldsymbol{w}(t),\{k_i(t)\}} \quad & \tilde{\eta}_{EE}=\lim_{T\to\infty} \frac{1}{T}\sum_{t=1}^{T}\eta_{EE}(t), \label{obj8}\\
	s.t.  \ \ \quad & \tilde{E}_h=\lim_{T\to\infty}\frac{1}{T}\sum_{t=1}^{T}E_h(t)\ge \tilde{E}_{th}, \tag{\ref{obj8}{a}}\label{obj8a}\\
	&\tilde{R}=\lim_{T\to\infty}\frac{1}{T} \sum_{t=1}^{T}R(t)\ge \tilde{R}_{th}, \tag{\ref{obj8}{b}}\label{obj8b}\\
	&||\boldsymbol{w}(t)||^2\leq P_{max}, \tag{\ref{obj8}{c}}\label{obj8c}\\ \notag
\end{align}
where \eqref{obj8a} and \eqref{obj8b} represent the long term constraints of achievable throughput and harvested energy, respectively. Since there is no long-term transmit power constraint in (P8), the DRL agent only needs to make the port selection decisions based on the CSI of each transmission frame, while the optimal beamforming vector can be obtained from (P7). According to \cite{10542705}, (P8) can be modeled as a constrained Markov decision process (CDMP), which is reformulated as

 \begin{align}
	\text{(P9)}:&\max_{\pi} \quad  \mathbf{E}_{\pi}\left[ (1-\beta)\sum_{t=1}^{\infty}\beta^{t-1}\eta_{EE}(s_t,a_t)\right], \label{obj9}\\
	s.t.  &\ \mathbf{E}_{\pi}\left[ (1-\beta)\sum_{t=1}^{\infty}\beta^{t-1}(E_h(s_t,a_t)-\tilde{E}_{th})\right]\ge 0, \tag{\ref{obj9}{a}}\label{obj9a}\\
	&\ \mathbf{E}_{\pi}\left[ (1-\beta)\sum_{t=1}^{\infty}\beta^{t-1}(R(s_t,a_t)-\tilde{R}_{th})\right]\ge 0, \tag{\ref{obj9}{b}}\label{obj9b}\notag
\end{align}
where $\eta_{EE}(s_t,a_t)$, $E_h(s_t,a_t)$ and $R(s_t,a_t)$ denote the instantaneous WET efficiency, harvested energy and achievable throughput by taking action $a_t$ at state $s_t$, respectively. $\beta$ is the discount factor. In (P9), the expected long term discount WET efficiency is maximized by searching for the optimal port selection policy $\pi^*$, while the constraints of expected long term discount throughput and harvested energy are guaranteed.  However, (P9) is hard to be directly solved by DRL algorithm due to the expected long-term discount constraints. Therefore, we transform (P9) into multi-objective MDP problem and use zero-sum Markov Bandit game to solve it \cite{RL-cdmp}. Maximizing the objective of (P9) is equivalent to obtain the maximal value of $\delta$ by satisfying
\begin{align}
	\mathbf{E}_{\pi}\left[ (1-\beta)\sum_{t=1}^{\infty}\beta^{t-1}\eta_{EE}(s_t,a_t)\right]\ge\delta.
\end{align}
Therefore, (P9) can be reformulated as
 \begin{align}
	\text{(P10)}:& \max_{\pi} \ \delta \label{obj10}\\
	s.t. & \ \mathbf{E}_{\pi}\left[ (1-\beta)\sum_{t=1}^{\infty}\beta^{t-1}(\eta_{EE}(s_t,a_t)-\delta)\right]\ge0,\tag{\ref{obj10}{a}}\label{obj10a}\\
	 & \mathbf{E}_{\pi}\left[ (1-\beta)\sum_{t=1}^{\infty}\beta^{t-1}(E_h(s_t,a_t)-\tilde{E}_{th})\right]\ge 0, \tag{\ref{obj10}{b}}\label{obj10b}\\
	&\mathbf{E}_{\pi}\left[ (1-\beta)\sum_{t=1}^{\infty}\beta^{t-1}(R(s_t,a_t)-\tilde{R}_{th})\right]\ge 0. \tag{\ref{obj10}{c}}\label{obj10c}\notag
\end{align}
In (P10), the discounted WET efficiency constraint prevents the agent from prioritizing short-term energy gains, which could lead to excessive port  switching delay and future performance degradation. The discounted throughput constraint ensures that the average throughput stays above the required level, preventing the agent from overly favoring energy transfer at the cost of communication quality. These constraints influence the agent’s long-term decisions: a higher average energy target promotes the long-term WET efficiency, potentially sacrificing throughput, while a higher average throughput target prioritizes DR performance, which may reduce energy harvesting performance and WET efficiency of ER. The maximum $\delta$ can be found by exploiting bisection search, where a feasible policy $\pi$ can be found by satisfying the constraints \eqref{obj10a} to \eqref{obj10c}. When the objective value $\delta$ of (P10) is optimal, the corresponding policy $\pi$ is also the optimal policy. In order to solve (P10), we formulate the problem as zero-sum Markov-Bandit game under a fixed objective value $\delta$, where the DRL agent solves a Markov decision problem, while its opponents solve a Bandit optimization problem. The zero-sum Markov-Bandit game is defined by the tuple $(\mathcal{S},\mathcal{A},\mathcal{O},\mathcal{P},\mathcal{R})$, where $\mathcal{S}$ is the state space,  $\mathcal{A}$ is the action space, $\mathcal{O}=\{o_1,o_2,o_3\}$ denotes the opponents' space, which corresponds to the expected long-term discount constraints of (P10). $\mathcal{P}$ represents the transition probability function among different states and $\mathcal{R}:\mathcal{S}\times\mathcal{A}\to\mathcal{R}$ is the expected reward function. The zero-sum Markov-Bandit game is aimed to maximize the total discounted reward given by
\begin{align}
	V_{\pi}(s_t)&=\min_{o} \mathbf{E}_{\pi}\left[ Q(s_t,a_t,o)\right] \notag\\ &=\min_{o}\mathbf{E}_{\pi}\left[ \sum_{t=1}^{
	\infty}\beta^{t-1}R(s_t,a_t,o)\right],
\end{align}
where $R(s_t,a_t,o)$ denotes the reward function for taking action $a_t$ under current state $s_t$, which is corresponding to the expected long-term discount constraints in (P10). $Q(s_t,a_t,o)$ is the action value function representing the expected return obtained by taking action $a_t$ in state $s_t$ and then following policy $\pi$ thereafter. Note that the reward function $R(s_t,a_t,o)(o\in\mathcal{O})$ should be designed to reflect the satisfaction of each constraint and a larger $Q(s_t,a_t,o)$ means a better satisfaction of constraint $o$. The optimal policy $\pi^*$ of zero-sum Markov-Bandit game can be obtained  as
\begin{align}
\pi^*=\text{arg} \max_{\pi} V_{\pi}(s_t),\forall s_t\in\mathcal{S}.
\end{align}
If the state value function $ V_{\pi}(s_t)$ satisfies $ V_{\pi}(s_t)>0(\forall s_t\in\mathcal{S})$, $\pi^*$ is a feasible policy that satisfies all the long term constraints of (P10) and vice versa.

\subsection{DRL Design for Port Selection}
\begin{figure*}
	\centering
	\includegraphics[width=0.8\linewidth]{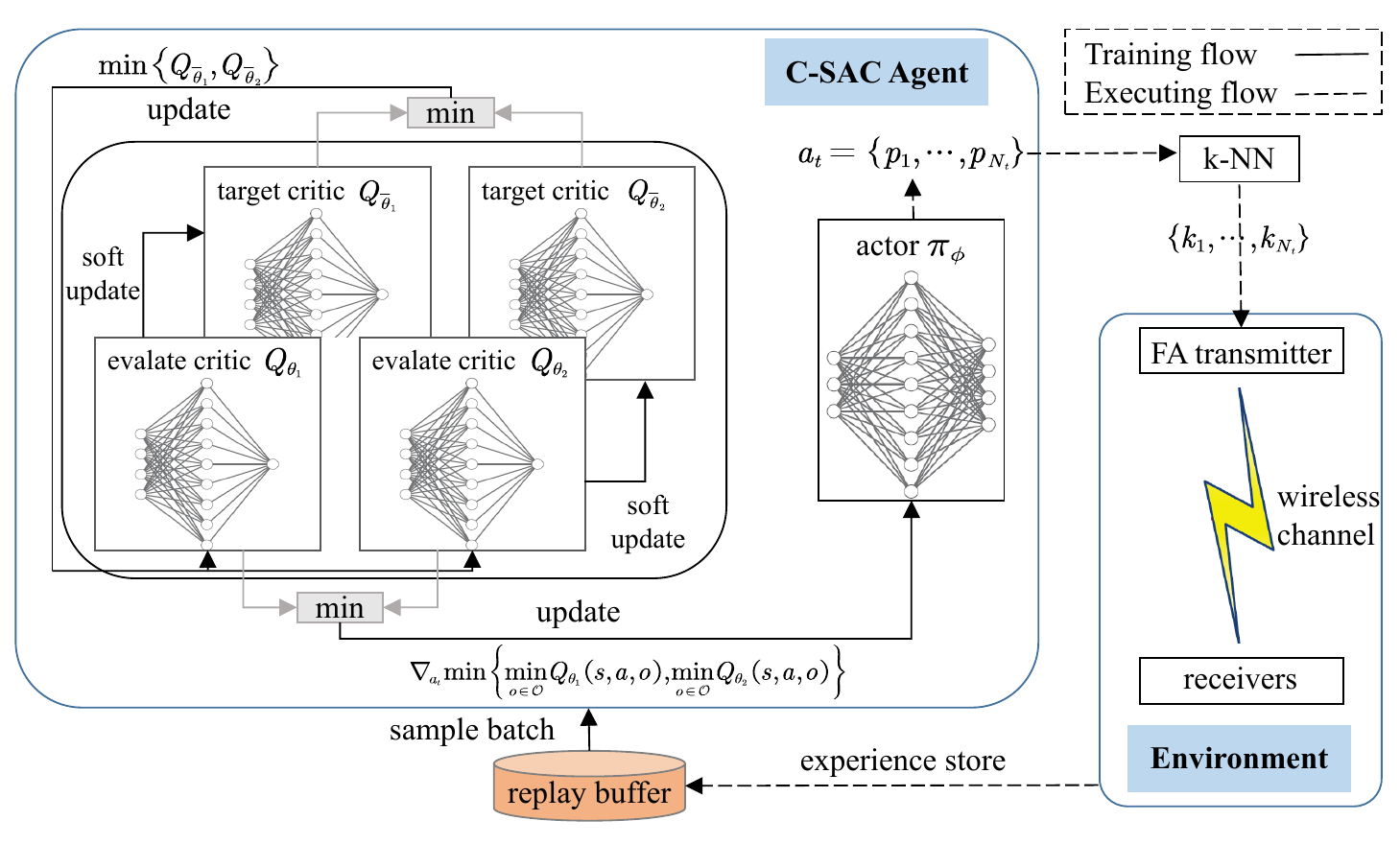}
	\caption{The framework of C-SAC algorithm}
	\label{fig:c-sac}
\end{figure*}
Due to the large number of fluid antenna ports, using discrete actions to describe port selection would lead to an excessively large action space (e.g. $N^{N_t}$). Therefore, we use continuous actions to denote the port selection and then use $k$-Nearest Neighbors ($k$-NN) algorithm to obtain the antenna position closet to the output actions. In deep reinforcement learning family, deep deterministic policy gradient (DDPG) algorithm is an off-policy actor-critic method designed for continuous action spaces, offering effective learning through the use of experience replay and target network. However, due to its sensitivity to hyperparameters and limited exploration capabilities, obtaining good results with DDPG  algorithm is usually a challenge in some environments \cite{10064035}. Soft actor critic (SAC) is the extend version of DDPG by introducing an entropy term in value function to enhance exploration, which can achieve more robust performance and stable learning than DDPG. Motivated by the pioneering works \cite{10542705}\cite{sac1}, we have reformulated the soft actor critic framework to explore feasible policies in the zero-sum Markov-Bandit game, resulting in a novel constrained SAC (C-SAC) algorithm.
\subsubsection{DRL definitions}
In our FAS assisted IDET system, the states, actions, reward functions and discount factor are defined as follows:

$\bullet$\textit{ States}: The state $s_t\in \mathcal{S}$ in $t$-th transmission frame is defined as the channel gain between transmitter and DR and ER because such information is highly related to both the WET and WDT performance. However, due to the large number of fluid antenna ports, using all port channel gains as the state would lead to a prohibitively high dimensional state space, making the DRL model difficult to train and converge. To address this, we exploit the strong spatial correlation among ports and uniformly sample the channel gains from a subset of them to form the state. In addition, the position of metallic droplet should also be incorporate into the state.  Therefore, the state can be expressed as
\begin{align}
	\boldsymbol{s}_t=[&|\boldsymbol{g}_{1,L_s}(t)|,\cdots,|\boldsymbol{g}_{N_{t},L_s}(t)|,|\boldsymbol{h}_{1,L_s}(t)|,\cdots,|\boldsymbol{h}_{N_{t},L_s}(t)|, \notag\\ &k_1(t),\cdots,k_{N_t}(t)],
\end{align}
where $|\boldsymbol{g}_{i,L_s}|$ and $|\boldsymbol{h}_{i,L_s}|$ denote the sampled channel gain from $i$-th fluid antenna to ER and DR, respectively. $L_s$ is the total number of samples.

$\bullet$\textit{Actions}: In the $t$-th transmission frame, the DRL agent needs to provide the port selection of each fluid antenna to the transmitter according to the instantaneous CSI and antenna location. Accordingly, the action is designed as a $N_t$ dimension vector $\boldsymbol{a}_t=[p_1,\cdots,p_{N_t}],\forall p_i \in[1,N]$, where $p_i$ denotes the $i$-th antenna position. Note that the outputs of DRL actor network are continuous actions, the transmitter needs to use $k$-NN algorithm to obtain the closest discrete position.

$\bullet$\textit{Reward function}:  In  zero-sum Markov-Bandit game, the reward function is comprised of the agent reward function and the opponents reward function, which correspond to the constraints \eqref{obj10a} to \eqref{obj10c} in (P10). Therefore, the reward function is defined as
\begin{align}\label{R}
	R(s_t,a_t,o)=\left\{
	\begin{aligned}
		&\eta_{EE}(\boldsymbol{s}_t,\boldsymbol{a}_t)-\overline{\delta},\qquad o=o_1,\\
		&E_h(\boldsymbol{s}_t,\boldsymbol{a}_t)-\tilde{E}_{th},\qquad o=o_2,\\
		&R(\boldsymbol{s}_t,\boldsymbol{a}_t)-\tilde{R}_{th},\qquad o=o_3,\\
	\end{aligned}
	\right.
\end{align}
where $o\in\mathcal{O}=\{o_1,o_2,o_3\}$ denote the opponent space of DRL agent, $\overline{\delta}$ is the target objective value of (P10).

$\bullet$\textit{Discount factor}:  The discount factor $\beta$ (typically close to 1) determines the relative importance of immediate rewards versus future rewards in the DRL training process. A larger discount factor $\beta$ endows the agent with greater foresight by emphasizing future rewards, while a lower discount factor $\beta$ makes the agent to prioritize the current rewards. In order to maintain the equivalence between (P8) and (P9), $\beta$ should be set close to 1, ensuring that the agent effectively captures the long-term dynamics of the wireless environment and mitigates the risk of converging to a local optimum.

\subsubsection{Framework and updating process of the C-SAC algorithm}
SAC is an off-policy reinforcement learning algorithm that leverages a generalized maximum entropy objective to derive robust stochastic policies. The classic SAC algorithm aims to maximize the expected reward with an entropy term $\mathcal{H}$ to obtain the optimal policy, which can be denoted as
\begin{align}
	\pi^*=\text{arg} \max_{\pi} \mathbf{E}_{(\boldsymbol{s}_t,\boldsymbol{a}_t)\sim\rho_{\pi}}\left[ \sum_{t=1}^{\infty}\beta^{t-1}\left( R(\boldsymbol{s}_t, \boldsymbol{a}_t) \right. \right. \notag\\
	\left. \left. + \alpha \mathcal{H}\left( \pi\left(\cdot|s_t\right) \right) \right) \right],
\end{align}
where $\alpha$ is temperature parameter to control the stochasticity of the policy by adjusting the importance of the entropy term;  $\mathcal{H}(\pi(\cdot|s_t))=\mathbf{E}_{a_t}\left[ -\log\pi(a_t|s_t)\right]$ is the policy entropy. Therefore, when the SAC algorithm is adopted to tackle the zero-sum Markov-Bandit game, the optimal policy $\bar{\pi}^*$  can be obtained via
\begin{align}
	\bar{\pi}^* = \text{arg} \max_{\pi} \min_{o} \mathbf{E}_{ (\boldsymbol{s}_t, \boldsymbol{a}_t) \sim \rho_{\pi} } \left[  \sum_{t=1}^{\infty} \beta^{t-1} \left( R(\boldsymbol{s}_t, \boldsymbol{a}_t, o) \right. \right. \notag\\
	\left. \left. + \alpha \mathcal{H}\left( \pi\left(\cdot|s_t\right) \right) \right) \right].
\end{align}

 According to the Bellman backup equation, the soft $Q$ function of opponent $o$ can be defined as
\begin{align}
	Q(\boldsymbol{s}_t,\boldsymbol{a}_t,o)=R(\boldsymbol{s}_t,\boldsymbol{a}_t,o)+\beta \mathbf{E}_{\boldsymbol{s}_{t+1}\sim\rho_{\pi}}\left[ V(\boldsymbol{s}_{t+1},o)\right],
\end{align}
 where
\begin{align}
	 V(\boldsymbol{s}_{t},o)=\mathbf{E}_{\boldsymbol{a}_{t}\sim\pi}\left[Q(\boldsymbol{s}_t,\boldsymbol{a}_t,o)-\alpha\log\pi(\boldsymbol{a}_t|\boldsymbol{s}_t)\right]
\end{align}
is the soft state value function. Therefore, the C-SAC agent is comprised of an actor network and two groups of critic networks to approximate the policy function and soft $Q$ function, as shown in Fig. \ref{fig:c-sac}. The actor network outputs the target position of each antenna and the critic networks evaluate the action-value pair $Q(\boldsymbol{s}_t,\boldsymbol{a}_t,o)$ based on the current state. In addition, two target critic networks are employed to alleviate the overestimation of $Q$ value and enhance learning stability.

In order to obtain the optimal port selection policy, the SAC  algorithm uses stochastic gradient descent (SGD) to update the deep neural network (DNN) at each training step. Two group of critic networks have the same  DNN structure, and each group of critic network consists of a critic evaluate network and a  critic target network.  These critic networks can be denoted as $Q_{\theta_1}$, $Q_{\bar{\theta}_1}$,  $Q_{\theta_2}$ and $Q_{\bar{\theta}_2}$, where $\theta_{n},\bar{\theta}_{n},n=\left\lbrace 1,2\right\rbrace $ are the DNN weights of critic evaluate networks and critic target networks, respectively. The parameters of each critic network are updated by minimizing the soft Bellman residual, which can be represented as
\begin{align}
	J_Q(\theta_n)=\mathbf{E}_{(\boldsymbol{s}_t,\boldsymbol{a}_t,o_i)\sim\mathcal{D}}\left[\frac{1}{2} (Q_{\theta_n}(\boldsymbol{s}_t,\boldsymbol{a}_t,o_i)-y)^2\right],
\end{align}
where $\mathcal{D}$ denotes a batch of experience items sampled from the replay buffer (i.e., the tuple $(\boldsymbol{s}_t,\boldsymbol{a}_t,o_i,r_t,\boldsymbol{s}_{t+1}\in\mathcal{D})$) and
\begin{align}
	y=r_t+\beta\mathbf{E}_{\boldsymbol{s}_{t+1}}\left[\min_{n\in\left\lbrace 1,2\right\rbrace} \left\lbrace Q_{\bar{\theta}_n}(\boldsymbol{s}_{t+1},\boldsymbol{a}_{t+1},o_i)\right\rbrace \right.\notag\\ \left.-\alpha\log\pi_{\phi}(\boldsymbol{a}_{t+1}|\boldsymbol{s}_{t+1})\right].
\end{align}
The stochastic gradients of Bellman residual can be calculated as follows
\begin{align}
	&\nabla_{\theta_n} J_Q(\theta_n)=\nabla Q_{\theta_n}(\boldsymbol{s}_t,\boldsymbol{a}_t,o_i)\left(  Q_{\theta_n}(\boldsymbol{s}_t,\boldsymbol{a}_t,o_i)- r_t \right.\notag\\
    &\left.-\beta\left[ \min_{n\in\left\lbrace 1,2\right\rbrace} \left\lbrace Q_{\bar{\theta}_n}(\boldsymbol{s}_{t+1},\boldsymbol{a}_{t+1},o_i)\right\rbrace-\alpha\log\pi_{\phi}(\boldsymbol{a}_{t+1}|\boldsymbol{s}_{t+1})\right] \right).
\end{align}
Therefore, the weights of critic evaluate networks are updated by $\theta_{n}(t)\gets\theta_{n}(t-1)-\lambda_Q	\nabla J_Q(\theta_n)$ with the learning rate $\lambda_Q$. After updating the evaluation networks'  parameters, the target networks' parameters are obtained through a soft update from the evaluation networks, which can be expressed as
\begin{align}
	\bar{\theta}_{n}\gets \mu \bar{\theta}_{n}+(1-\mu)\theta_{n},n=1,2,
\end{align}
where $\mu$ is the soft update factor.

In zero-sum Markov Bandit game, the actor network is updated by maximizing the minimum $Q$ value among all the opponents $o\in\mathcal{O}$, while in SAC algorithm, the actor network is updated by minimizing the Kullback-Leibler (KL) divergence. Therefore, the actor network in our C-SAC algorithm can be trained by minimizing the expected KL divergence with the minimum $Q$ value of all opponents \cite{sac2}:
\begin{align}
	J_{\pi}(\phi)=\mathbf{E}_{\boldsymbol{s}_{t}\sim\mathcal{D}} \left[ \mathbf{E}_{\boldsymbol{a}_{t}\sim\pi_{\phi}} \left[ \alpha\log\pi_{\phi}(\boldsymbol{a}_{t}|\boldsymbol{s}_{t})\right. \right.\notag\\
	\left. \left.-\min_{n\in\left\lbrace 1,2\right\rbrace} \left\lbrace \min_{o\in\mathcal{O}}Q_{{\theta}_n}(\boldsymbol{s}_{t},\boldsymbol{a}_{t},o)\right\rbrace\right] \right].
\end{align}
As the actor network parameters $\phi$ cannot be back propagated in a normal way if we directly generate samples by using $\pi_{\phi}(\boldsymbol{a}_{t}|\boldsymbol{s}_{t})$, it is thus convenient to apply the reparameterization trick instead. Hence, the port selection action can be generated by
\begin{align}
	\boldsymbol{a}_t=f_{\phi}(\epsilon_t;\boldsymbol{s}_t),
\end{align}
where $\epsilon_t$ is a noise vector and $f_{\phi}$ is fixed distribution function. Therefore, the gradient of KL divergence can be approximate with
\begin{align}
	\nabla_{\phi} J_\pi(\phi)\approx\nabla_{\phi}\alpha\log\pi_{\phi}(\boldsymbol{a}_{t}|\boldsymbol{s}_{t})+\left(\nabla_{\boldsymbol{a}_t}\alpha\log\pi_{\phi}(\boldsymbol{a}_{t}|\boldsymbol{s}_{t})\right.\notag\\
	\left.-\nabla_{\boldsymbol{a}_t}\min_{n\in\left\lbrace 1,2\right\rbrace} \left\lbrace \min_{o\in\mathcal{O}}Q_{{\theta}_n}(\boldsymbol{s}_{t},\boldsymbol{a}_{t},o)\right\rbrace \right). \nabla_{\phi}f_{\phi}(\epsilon_t;\boldsymbol{s}_t)
\end{align}
Note that $\nabla_{\boldsymbol{a}_t}\min_{n\in\left\lbrace 1,2\right\rbrace} \left\lbrace \min_{o\in\mathcal{O}}Q_{{\theta}_n}(\boldsymbol{s}_{t},\boldsymbol{a}_{t},o)\right\rbrace$ is calculated with the latest DNN weights $\theta_{n}(t)$ of the critic evaluate networks. The parameter of actor network is then updated by $\phi(t)\gets\phi(t-1)-\lambda_{\pi}\nabla_{\phi} J_\pi(\phi)$ with learning rate $\lambda_{\pi}$.
As a fixed temperature factor $\alpha$ may cause the training process instability due to the imbalance between exploration and exploitation. Therefore, an automatic entropy adjustment for maximum entropy RL is necessary. The temperature factor $\alpha$  can be trained by minimizing the loss of temperature, which can be expressed as
\begin{align}
	J(\alpha)=\mathbf{E}_{\boldsymbol{s}_t\sim\mathcal{D}}\mathbf{E}_{\boldsymbol{a}_t\sim\pi_{\phi}}\left[-\alpha\log\pi_{\phi}(\boldsymbol{a}_{t}|\boldsymbol{s}_{t})-\alpha\mathcal{H}_0\right].
\end{align}
where $\mathcal{H}_0$ is the target entropy. The details of C-SAC algorithm for port selection are summarized in Algorithm 2.

\begin{algorithm}[t]
	\renewcommand{\algorithmicrequire}{\textbf{Require:}}
	\renewcommand{\algorithmicensure}{\textbf{\textsc{Repeat}:}}
	\caption{The C-SAC Algorithm for searching feasible police of port selection}
	\label{power}
	\begin{algorithmic}[1]
		\REQUIRE   Given a WET efficiency target objective value $\bar{\delta}$
		
		\STATE Initialize the discount factor $\beta$, batch size $B_s$, replay buffer length $L_{buffer}$, learning rate $\lambda_{\alpha}$, $\lambda_Q$ and $\lambda_{\pi}$, soft update factor $\mu$, maximum training epoch number $T_{max}$.
		\STATE Initialize the actor network parameter $\phi$, evaluate critic networks parameter $\theta_1$ and $\theta_2$, target critic networks parameter  $\bar{\theta}_1\gets{\theta}_1$ and $\bar{\theta}_2\gets{\theta}_2$
		
		\FOR {t=1 to $T_{max}+5B_s$ transmission frames}
		\STATE  Input current state $s_t$ into actor network and get continuous position action $\boldsymbol{a}_t\sim\pi_{\phi}(\cdot|\boldsymbol{s}_t)$
		\STATE  {Select the discrete position closet to the continuous position action as transmitting port, solve (P7) to get the beamforming vector, calculate the WET efficiency reward $r_{t,1}=r_t(\boldsymbol{s}_t,\boldsymbol{a}_t,o_1)$, the achievable rate $r_{t,2}=r_t(\boldsymbol{s}_t,\boldsymbol{a}_t,o_2)$ and the harvested energy reward $r_{t,3}=r_t(\boldsymbol{s}_t,\boldsymbol{a}_t,o_3)$ according to Eq. \eqref{R}; get next state $s_{t+1}$}.
		\STATE  Store the transitions $(\boldsymbol{s}_t,\boldsymbol{a}_t,o,r_{t,o},s_{t+1})$ for all opponents $o\in\mathcal{O}$ into experience replay buffer.
		\IF  {$t>5B_s$}
		\STATE  Take $B_s$ samples from the replay buffer
		\STATE  Update the evaluation critic networks $\theta_{n}(t)\gets\theta_{n}(t-1)-\lambda_Q	\nabla J_Q(\theta_n),n={1,2} $
		\STATE  Update the actor networks $\phi(t)\gets\phi(t-1)-\lambda_{\pi}\nabla_{\phi} J_\pi(\phi) $
		\STATE   Update the temperature factor $\alpha(t)\gets\alpha(t-1)-\lambda_{\alpha}\nabla_{\alpha} J(\alpha)$
		\STATE    Update the target critic networks $\bar{\theta}_{n}\gets \mu \bar{\theta}_{n}+(1-\mu)\theta_{n},n={1,2}$
		\ENDIF
		\ENDFOR
		\RETURN The optimal actor network $\pi_{\phi}(a_t|s_t)$ under WET efficiency target objective value $\bar{\delta}$
	\end{algorithmic}
\end{algorithm}

\subsection{Complexity analysis}
The computational complexity of the C-SAC algorithm primarily arises from two aspects: the training complexity and the execution complexity \cite{10129152}. During the training phase, the gradient descent complexity for both the actor network and the critic network can be expressed as \(\mathcal{O}\left(B_s \left( \sum_{k=0}^{K-1} n_k^{\pi} n_{k+1}^{\pi} + \sum_{j=0}^{J-1} n_j^Q n_{j+1}^Q \right)\right)\), where \(B_s\) is the batch size, \(K\) and \(J\) represent the number of fully connected layers in the actor and critic networks, respectively. \(n_k^{\pi}\) and \(n_j^Q\) denote the number of neurons in the \(k\)-th layer of the actor network and the \(j\)-th layer of the critic network, respectively, with \(k=0\) and \(j=0\) referring to the input layers. Therefore, the overall complexity for training the C-SAC algorithm is \(\mathcal{O}\left(T_{max} B_s \left( \sum_{k=0}^{K-1} n_k^{\pi} n_{k+1}^{\pi} + \sum_{j=0}^{J-1} n_j^Q n_{j+1}^Q \right)\right)\), where \(T_{max}\) is the maximum number of training epochs. During the execution phase, only the trained actor network is required, and thus the computational complexity at each transmission frame corresponds to a forward pass through the actor network, which can be expressed as \(\mathcal{O}\left( \sum_{k=0}^{K-1} n_k^{\pi} n_{k+1}^{\pi} \right)\).

\begin{table}
	\centering
	\caption{FAS-IDET System Parameters setting}
	\label{tab0}
	\scalebox{0.7}{
	\begin{tabular}{|c|c|c|c|}
		\hline
		\makecell[c]{Parameter }&\makecell[c]{value } &\makecell[c]{Parameter}&\makecell[c]{value}\\
		\hline
		Number of FAs ($N_t$)&4 & Dimension ratio $(D/L)$& 5\\
	%	\hline
		Number of ports ($N$)&50 &Coherence time $(T)$& 50 ms\\
	%	\hline
		Wavelength ($\lambda$) &6cm &Noise variance $(\sigma^2)$& $10^{-8}$W\\
	%	\hline
		Scaling constant ($W$)&1 &Maximal transmission power ($P_{max}$) &1W\\
	%	\hline
		Initial charge $(q)$ &0.07V&Energy conversion efficiency $\eta$& 0.8 \\
	%	\hline
		Viscosity $(\mu)$ &0.002&Minimal harvested energy ($E_{th}$)& $5\times10^{-6}$\\
	%	\hline
		Slack penalty $p$ &1000&Circuit power&0.05W\\
		\hline
	\end{tabular}}
\end{table}
\section{Simulation Results}\label{section:sim}
\begin{table*}
	\centering
	\caption{C-SAC  Hyper Parameters}
	\label{tab1}
	\scalebox{0.7}{
		\begin{tabular}{|c|c|c|c|}
			\hline
			\makecell[c]{Parameter }&\makecell[c]{value } &\makecell[c]{Parameter}&\makecell[c]{value}\\
			\hline
			Batch size($B_s$)&256 & Target entropy $(\mathcal{H}_0)$& -$N_t$\\
			%	\hline
			Replay buffer length ($L_{buffer}$)&20000 &Average achievable data rate constraint $(\tilde{R}_{th})$& $8$bit/s/Hz\\
			%	\hline
			Discount factor  ($\beta$) &0.96 & Average harvested energy constraint $(\tilde{E}_{th})$& 2e-5\\
			%	\hline
			Maximum Training epoch ($T_{max}$)&12000 &Number of samples ($L_s$) &6\\
			%	\hline
			Actor network learning rate $(\lambda_{\phi})$ &1e-4 &Maximum WET efficiency objective value  ($\delta_{max}$)& 6e-4 \\
			%	\hline
			Critic network learning rate $(\lambda_{Q})$ &1e-3&Minimal WET efficiency objective value ($\delta_{min}$)& 3e-4\\
			%	\hline
			Temperature learning rate $(\lambda_{\alpha})$ &2e-4&Bisection searching accuracy&1e-5\\
			\hline
	\end{tabular}}
\end{table*}

In this section, we evaluate both the short term and long term WET efficiency of multiple FAs assisted IDET system by using simulation results. The distance from transmitter to DR and ER are set to be 10 meters and 3 meters, respectively. The signal attenuation at a reference distance of 1 meter is set as 30 dB. The path-loss factor of the both links are set to be 2.2. The whole model of our C-SAC algorithm is built on the platform of Pytorch 2.5.1 in Python 3.9. Both of the actor network and the critic networks have the dense layer 512$\times$256. The ReLU function is used for the hidden layers of both networks, while the Tanh and Linear functions are applied to the output layers of the actor and critic networks, respectively. The spatial and temporal correlation channel data is generated by Eq. \eqref{spa} and  Eq. \eqref{tem}. The temporal correlation factor is set as 0.98. Without specific statement, other default parameters about FAS assisted IDET system and and C-SAC algorithm are provided in Table \ref{tab0} and in Table \ref{tab1}. Note that those parameters are set for the sake of presentation. Using different parameters will lead to a shifted system performance, but with the same observations and conclusions.

\subsection{Evaluation of short term performance}
To evaluate the effectiveness of our proposed scheme, three benchmark schemes are considered in this section:
\begin{itemize}
	\item \textbf{Energy harvesting oriented scheme:} The harvested energy at the ER is maximized by jointly optimizing the beamforming vector and port selection, subject to the achievable throughput constraint. The optimization problem can be solved by using our proposed AO algorithm. This scheme reflects an energy-centric design that prioritizes energy transfer without explicitly considering overall WET efficiency.
	\item \textbf{Delay-insensitive scheme:} The WET efficiency is maximized by jointly optimizing the beamforming vector and port selection under the idealized assumption that port switching delay and power consumption are negligible, consistent with the common assumption in existing studies \cite{10506795,liao2025}. Note that while these costs are ignored during optimization, they are still accounted for during performance evaluation.
	\item \textbf{Fixed-position antenna (FPA) scheme:}  In this scheme, the fluid antennas at the transmitter are replaced by conventional fixed-position antennas (FPA). Thus, port switching delay and power consumption are irrelevant. The WET efficiency is maximized by optimizing the beamforming vector under both throughput and harvested energy constraints. The Dinkelbach's algorithm and SDR approach are adopted to obtain the optimal beamforming vector.
\end{itemize}

\begin{figure}[t]
	\centering
	\includegraphics[width=0.9\linewidth]{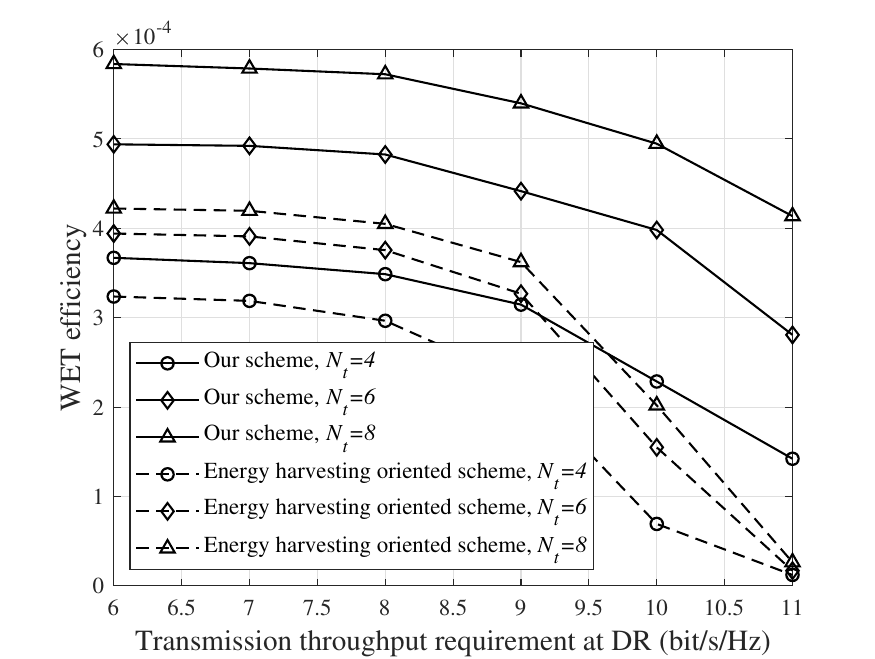}
	\caption{WET efficiency versus transmission throughput requirement at DR with different number of FAs $N_t$.}
	\label{fig:WET_1}
\end{figure}
Fig. \ref{fig:WET_1} depicts that  the trade-off between WET efficiency and the transmission throughput requirement at DR with different number of FAs. Meanwhile, the WET efficiency is compared between our proposed scheme and the energy harvesting oriented scheme. The voltage difference $\Delta\phi$ and moving power $P_u$ are set as 0.2 V and 2 W, respectively. Observe from Fig. \ref{fig:WET_1} that the WET efficiency decreases with the transmission throughput requirement at DR.  This is because under a higher throughput requirement, the optimization process prioritizes improving the signal quality at the DR to satisfy the stricter data rate constraint. As a result, the beamforming vector and port selection are steered more towards the DR, which may cause the transmission beam to become less favorable for the ER, thereby reducing the effective harvested energy. In addition, with the increment of transmitting FAs, the WET efficiency becomes higher. This is due to the additional antennas providing higher degrees of freedom, enabling more efficient wireless energy transfer. Moreover, the WET efficiency of our proposed scheme is much better than that of benchmark, which demonstrates that our proposed scheme can provide more  efficient integrate data and energy transfer.

\begin{figure}[t]
	\centering
	\includegraphics[width=0.9\linewidth]{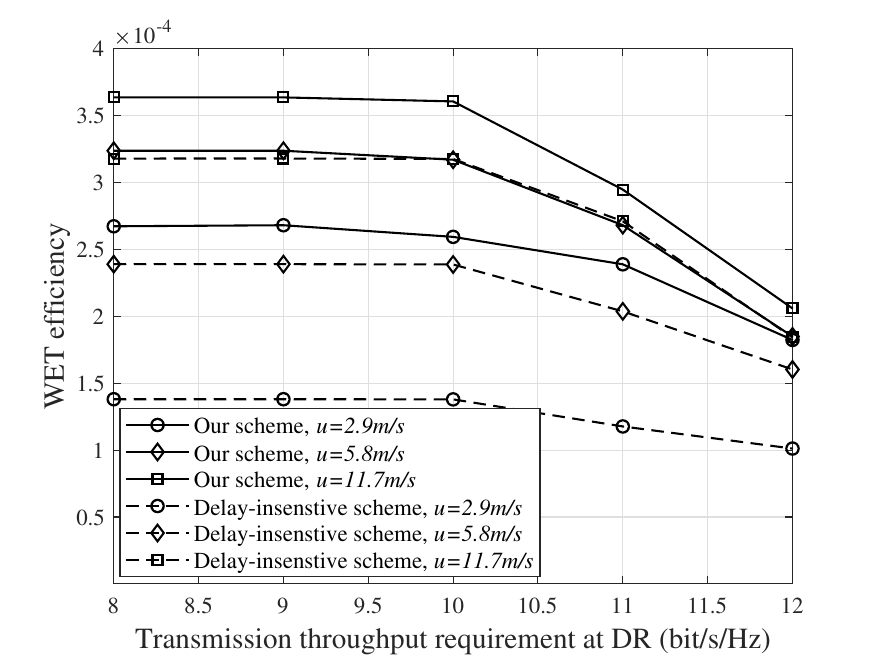}
	\caption{WET efficiency versus transmission throughput requirement at DR with different average velocity $u$ of droplet.}
	\label{fig:WET_2}
\end{figure}
Fig. \ref{fig:WET_2} compares the WET efficiency of our scheme and the delay-insensitive scheme under different average velocity $u$ of FA.
 \begin{comment}
		Note that both our proposed scheme and the delay-insensitive scheme perform port selection and beamforming design based on the estimated CSI prior to data transmission. Therefore, the port switching delay and power consumption are inherent in both schemes. However, in the delay-insensitive scheme, the port switching delay and energy consumption are not taken into account during the port selection and beamforming optimization process. The beamforming vector and port selection are jointly optimized in each transmission frame with the objective of maximizing the WET efficiency, while satisfying the achievable throughput constraint at the DR. The optimization process follows the same AO framework as our proposed method but assumes instantaneous port switching with zero energy cost. 
\end{comment}
	Observe from Fig. \ref{fig:WET_2} that the WET efficiency of our proposed scheme consistently outperforms that of the delay-insensitive scheme. This is because that the delay-insensitive scheme assumes that the FA movement is instantaneous and incurs no switching delay or energy consumption during the optimization process, but the actual system evaluation still accounts for these physical effects. Since the delay-insensitive scheme ignores the cost of FA movement in decision-making, it tends to generate aggressive port switching with large movement distances. This results in significant delays and energy consumption in practice, ultimately degrading the real WET efficiency. In contrast, our proposed scheme explicitly incorporates the port switching delay and energy consumption into the optimization, effectively reducing unnecessary movement and achieving higher practical WET efficiency. Furthermore, as the average droplet velocity increases, both the port switching delay and movement energy consumption decrease for both schemes, leading to improved WET efficiency. The performance gap between the two schemes narrows at higher velocities because the switching cost becomes less significant. However, even at high velocities, our proposed scheme consistently outperforms the delay-insensitive scheme by achieving more energy-aware port selection. The performance gain highlights the novelty of our work, as we are the first to explicitly incorporate the port switching delay and energy consumption of FAs into the optimization process. 

\begin{figure}[t]
	\centering
	\includegraphics[width=0.9\linewidth]{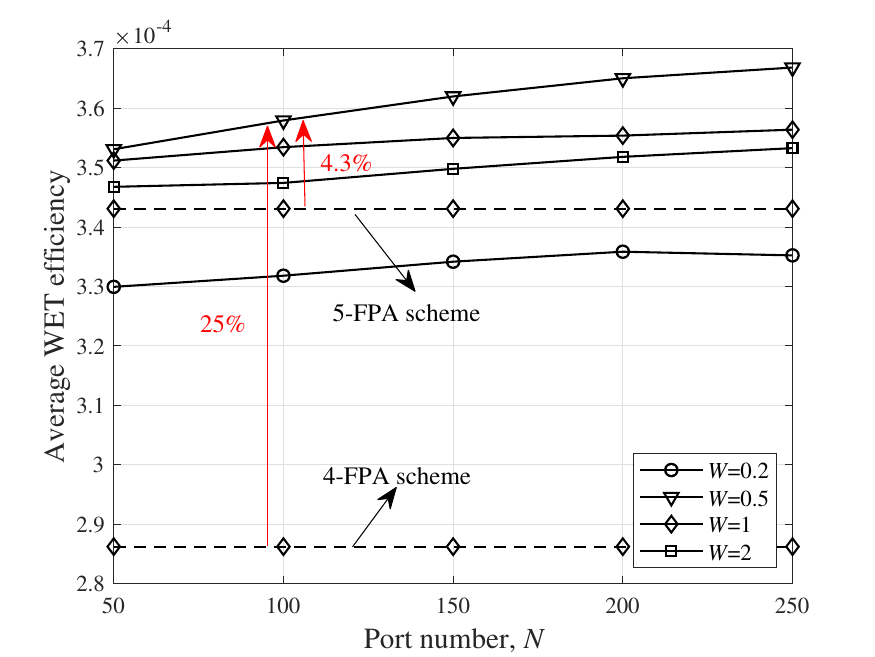}
	\caption{Average WET efficiency versus port number $N$ with different scaling antenna constant $W$.}
	\label{fig:WET_3}
\end{figure}

Fig. \ref{fig:WET_3} evaluates the average WET efficiency versus port number $N$ with different scaling constant $W$. In addition, the average WET efficiency is compared with that of FPA scheme. The achievable throughput  threshold is set as 7 bit/s/Hz.  As is shown in Fig. \ref{fig:WET_3}, when the FA has a port number $N=100$ and an antenna size $W=0.5$, the average WET efficiency of the FAS-assisted IDET system is 25\% higher than that of its 4-antenna FPA counterpart and 4.3\% higher than that of 5-antenna FPA system. This demonstrates that the FAS can achieve higher energy transfer efficiency with fewer antennas in IDET systems compared to conventional FPAs. This improvement is attributed to the additional spatial degrees of freedom provided by the FAS, which flexibly adjusts antenna positions to optimize energy transmission. In addition, the average WET efficiency increases with the port number, since a higher port number can capture more CSI from different positions, which results in a better port selection. However, different from the previous analysis in \cite{10506795} that a larger antenna size $W$ results in a better IDET performance, the average WET efficiency increases with $W$ initially and then decreases. This is because, although a larger antenna size reduces channel correlation, potentially leading to better channel quality, the delay and energy consumption will also increase accordingly. Therefore, selecting an appropriate antenna size is also crucial for improving the IDET performance when the switching delay and energy consumption of FA movement cannot be ignored.

\subsection{Evaluation of long term performance}
\begin{figure*}[t]
	\centering
	\includegraphics[width=0.8\linewidth]{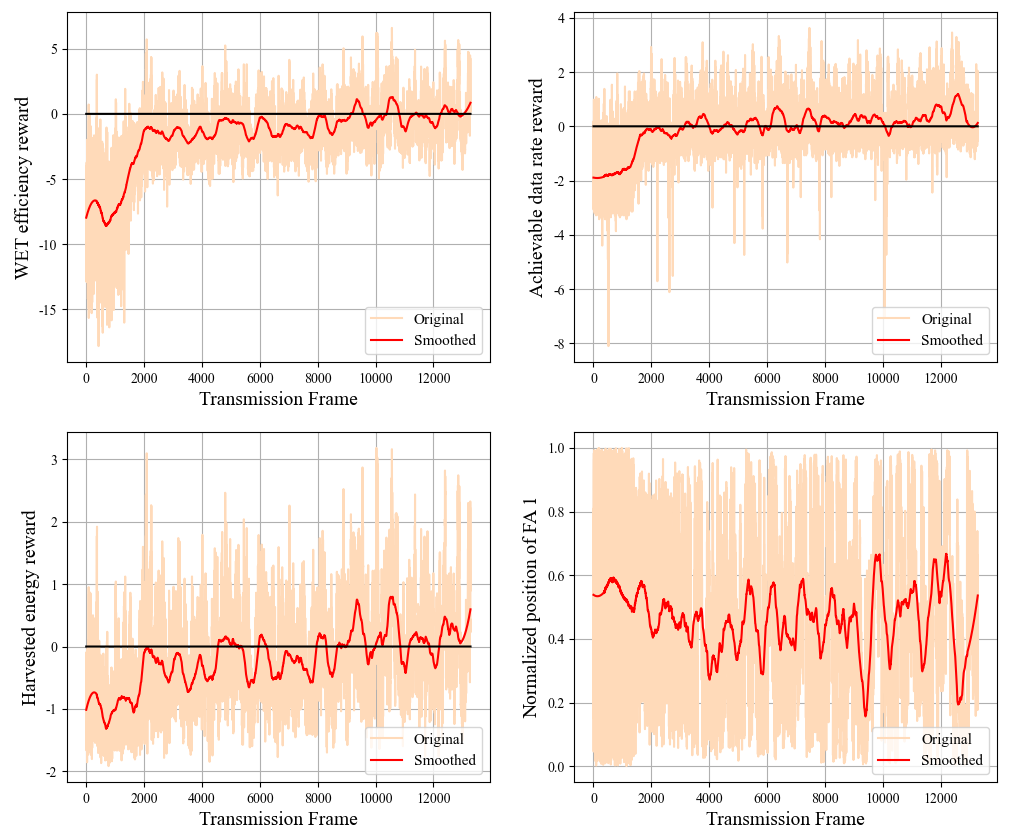}
	\caption{Convergence evaluation of C-SAC algorithm.}
	\label{fig:WET_4}
\end{figure*}
In this section, we evaluate the performance of C-SAC algorithm based on Python. The C-SAC agent interacts with a simulated time-varying wireless environment over 12,000 consecutive transmission frames, each corresponding to one channel coherence interval. The agent continuously updates its policy during these frames, capturing the dynamics of realistic online learning deployment. The voltage difference $\Delta\phi$ and moving power $P_u$ are set as 0.1 V and 2 W, respectively. In Fig. \ref{fig:WET_4}, we evaluate the convergence of our proposed C-SAC algorithm, where the WET efficiency reward, achievable throughout reward, harvested energy reward as well as one FA's normalized position are depicted in each transmission frame. The original curves represent the exact performance of each transmission frame, while the smoothed curves are obtained using Savitzky-Golay (SG) filter to show trends clearly \cite{9521788}. During the first 1280 transmission frames, the agent collects experience using the actor network with initial weights. In the subsequent 12000 transmission frames, the actor network and critic networks are trained by using the sampled experience stored in the replay buffer. Observe from Fig. \ref{fig:WET_4} at the beginning of 8000 transmission frames, the smoothed reward of WET efficiency and harvested energy are always lower than zero, which indicates that the long term WET efficiency constraint and harvested energy constraint are not satisfied. After the 8000-th transmission frame, all smoothed rewards vary around the zero line, which demonstrates our C-SAC algorithm converges. In addition, the normalized position of FA becomes more optimized as the network updates.

\begin{figure}[t]
	\centering
	\includegraphics[width=0.9\linewidth]{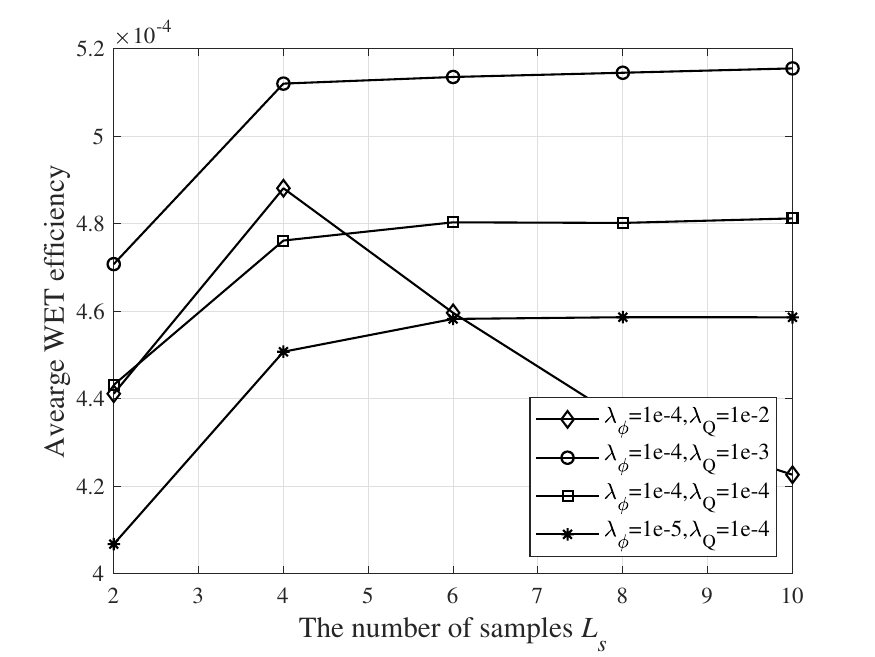}
	\caption{Average WET efficiency versus the number of samples $L_s$ with different learning rate.}
	\label{fig:WET_5}
\end{figure}

Fig. \ref{fig:WET_5} depicts the average WET efficiency versus the number of samples $L_s$ with different learning rate of critic networks. The average WET efficiency is calculated as the mean value over 1000 transmission frames by executing the port selection policy provided by the well-trained C-SAC agent. Observe from Fig. \ref{fig:WET_5} that with the increasing number of samples, the average WET efficiency is improved, except when the learning rate of critic network is 1e-2, which indicates that inadequate  sampling of FA channel may prevent the agent from capturing key positions of the FA,  leading to the output of a poor port selection policy. When the number of samples reaches 6, the average WET efficiency almost approaches its optimal value. However, when the learning rate of critic network is set to 1e-2, the average WET efficiency first increases and then decreases with the number of samples. This is because a larger learning rate and sample number make the model harder to converge, resulting in a decline in average WET efficiency. In addition, when the critic networks have a learning rate of 1e-3, the average WET efficiency outperforms other learning rates, since a smaller learning rate  is prone to lead the model to converge to a local optimum. Therefore, we set the learning rate of 1e-3 as the optimal learning rate of critic networks.

\begin{figure}[t]
	\centering
	\includegraphics[width=0.9\linewidth]{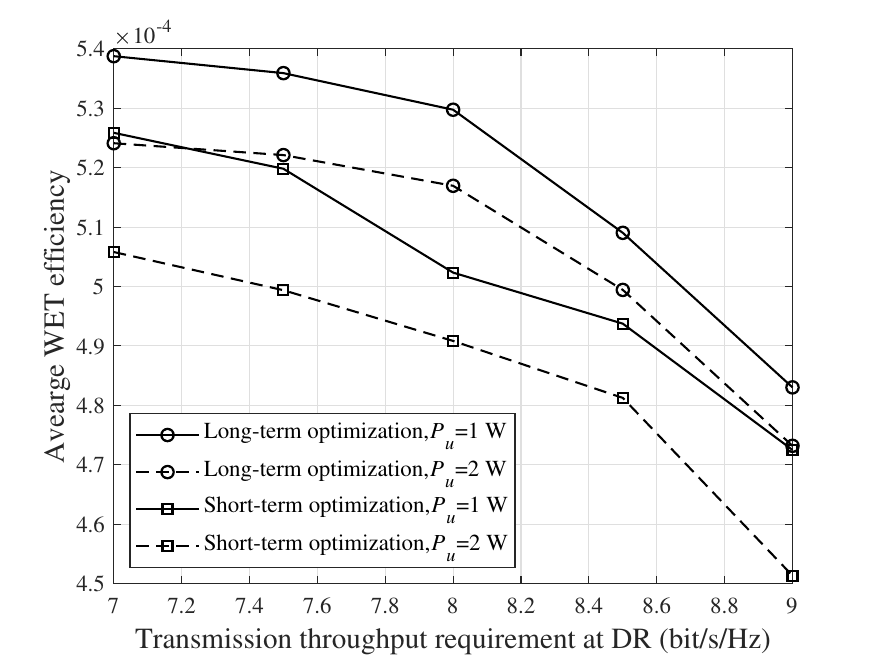}
	\caption{Average WET efficiency versus transmission throughput requirement at DR with different moving power $P_u$.}
	\label{fig:WET_6}
\end{figure}
Fig. \ref{fig:WET_6} compares the average WET efficiency performance between long-term optimization and short-term optimization under different moving power $P_u$. Observe from Fig. \ref{fig:WET_6} that the average WET efficiency of both optimization schemes decrease with the transmission throughput requirement, which reveals the trade-off between WDT performance and WET performance. In addition, compared to short-term optimization, the long-term optimization based on C-SAC algorithm achieves a better average WET efficiency. This is because the short-term optimization focuses on achieving the best instant WET efficiency in each transmission frame, which may result in lower WET efficiency in subsequent transmission frames. In contrast, long-term optimization leverages the C-SAC agent to capture the time-varying characteristics of the channel, enabling more proactive and forward-looking port selection decisions. Moreover, a larger moving power of FA results in a lower WET efficiency for both long-term optimization and short-term optimization, primarily due to the elevated energy consumption associated with higher FA movement.

\section{Conclusion}\label{section:conclusion}
In this paper, we investigated a fluid antenna-assisted IDET system while accounting for port switching delay and energy consumption. The beamforming vector and port selection are jointly optimized to maximize the WET efficiency from the short-term and long-term perspective, subject to the constraints of achievable throughput and energy harvesting amount. In the short-term WET efficiency optimization, the suboptimal solutions are obtained by alternately optimizing the port selection and beamforming vector using FPP-SCA and SDR. In the long-term WET efficiency optimization, a novel C-SAC algorithm is proposed to find the feasible policy for the transformed constraint satisfaction problem. Simulation results demonstrate that the short-term WET efficiency of our proposed scheme outperforms that of the energy harvesting oriented scheme as well as delay-insensitive scheme. Moreover, although the DRL training requires a relatively long convergence time, this cost is negligible compared to the typical operational lifetime of low-power wireless devices such as IoT sensors. The training overhead is acceptable and can be amortized over time. In addition, our results show that the converged DRL policy achieves significantly higher WET efficiency than short-term optimization schemes, highlighting its effectiveness for long-term energy management. This makes the proposed DRL-based solution well-suited for sustainable deployment in FAS assisted IDET system.

\bibliography{Reference}

\begin{IEEEbiography}[{\includegraphics[width=1in,height=1.25in,clip,keepaspectratio]{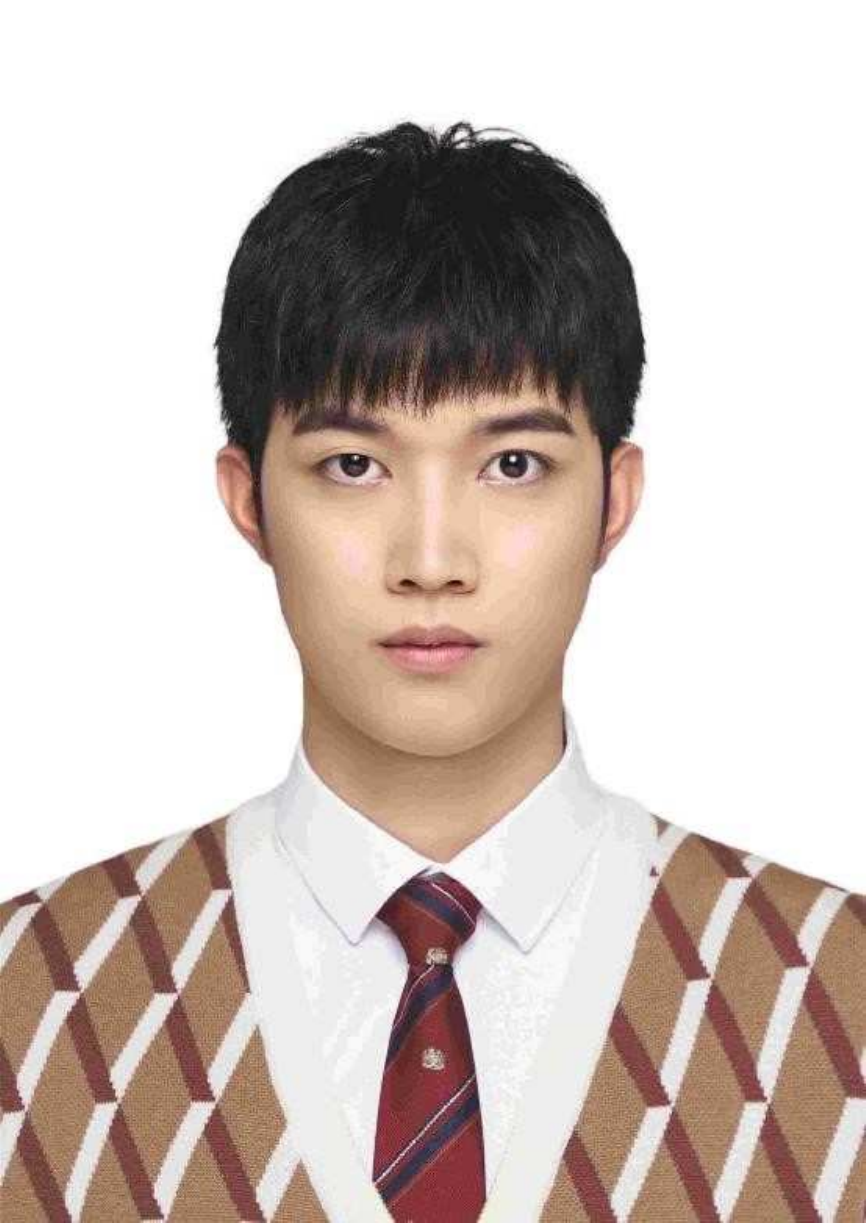}}]{Long Zhang }received the B.S. degree from Chongqing University in 2022. He is currently pursuing PHD in the School of Information and Communication Engineering, University of Electronic Science and Technology of China, Chengdu, China. His research interests include wireless communications, integrated data and energy transfer, fluid antenna systems.
\end{IEEEbiography}

\begin{IEEEbiography}[{\includegraphics[width=1in,height=1.25in,clip,keepaspectratio]{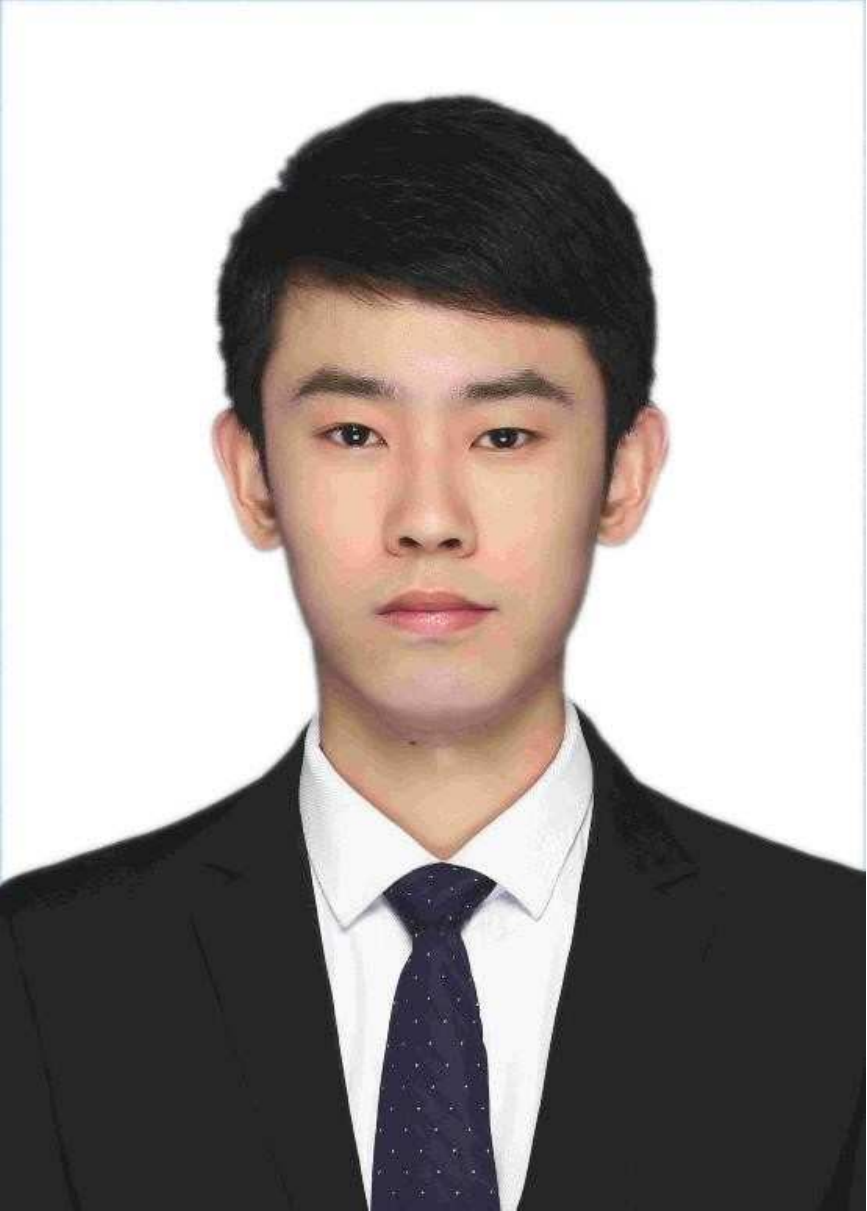}}]{Yizhe Zhao}[S'16, M'21] received the PhD in 2021 in School of Information and Communication Engineering from University of Electronic Science and Technology of China (UESTC), where he is currently an associate professor. He has been a visiting researcher with the Department of Electrical and Computer Engineering, University of California, Davis, USA. He is a member of IEEE and a senior member of China Institute of Communications. He is selected in Young Elite Scientists Sponsorship Program by China Association for Science and Technology (CAST). He serves for China Communications and Journal of Communications and Information Networks (JCIN) as the Guest Editor, and is also a TPC member of several prestigious IEEE conferences, such as IEEE ICC, Globecom. He was the recipient of IEEE CSE Best Paper Award in 2023. His research interests include modulation and coding design, integrated data and energy transfer, fluid antenna systems.
\end{IEEEbiography}

\begin{IEEEbiography}[{\includegraphics[width=1in,height=1.25in,clip,keepaspectratio]{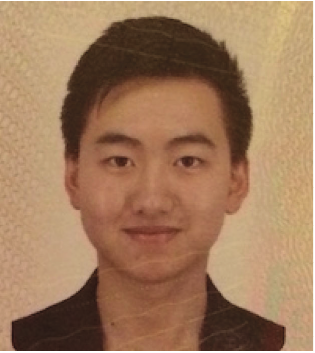}}]{Halvin Yang}
 received his B.Eng. degree in Electronic and Electrical Engineering from Imperial College London, U.K., in 2020, and his Ph.D. degree in Electronic and Electrical Engineering from University College London (UCL), U.K., in 2025. From April 2024 to September 2025, he was a Postdoctoral Research Associate with the Wolfson School of Mechanical, Electrical and Manufacturing Engineering, Loughborough University, U.K., affiliated with the EPSRC 6G flagship TITAN project. In October 2025, he joined Imperial College London, U.K., as a Research Fellow. His research interests include fluid antenna systems, orthogonal time frequency space (OTFS) modulation, non-terrestrial networks, and artificial intelligence for wireless communications. He has authored and coauthored papers in many IEEE Journals (For example, IEEE JSAC, IEEE TWC and IEEE ComMag), and his work has received multiple Best Paper Awards at international conferences.
\end{IEEEbiography}

\begin{IEEEbiography}[{\includegraphics[width=1in,height=1.25in,clip,keepaspectratio]{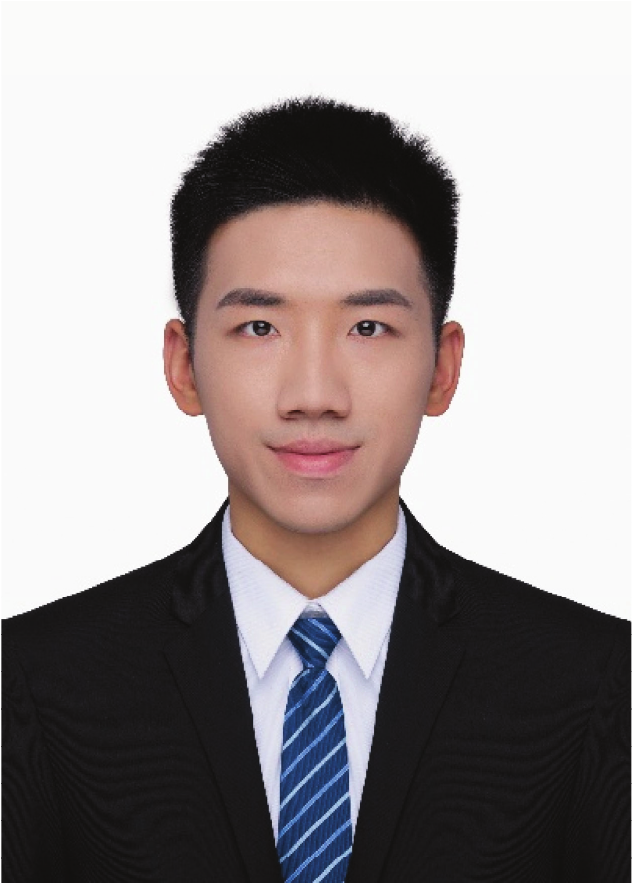}}]{Guangming Liang}
received the B.Eng. degree from Sun Yat-sen University (SYSU), Guangzhou, China, in 2021 and the M.Sc. degree from University of Electronic Science and Technology of China (UESTC), Chengdu, China, in 2024. He is currently pursuing the Ph.D. degree with the School of Computing Science, University of Glasgow (UofG), Glasgow, U.K. His current research interests include wireless communications, wireless networking, and artificial intelligence (AI) for communications. He is a recipient of the UofG Graduate School Scholarship (i.e., an International Full Scholarship) for his Ph.D. engagement. He serves as a TPC Member of several prestigious IEEE conferences, such as IEEE ICC, Globecom and VTC.
\end{IEEEbiography}

\begin{IEEEbiography}[{\includegraphics[width=1in,height=1.25in,clip,keepaspectratio]{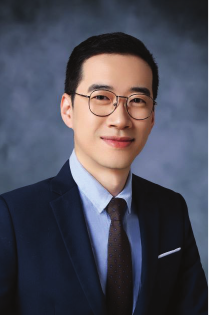}}]{Jie Hu}
	[S'11, M'16, SM'21] received his B.Eng. and M.Sc. degrees from Beijing University of Posts and Telecommunications, China, in 2008 and 2011, respectively, and received the Ph.D. degree from the School of Electronics and Computer Science, University of Southampton, U.K., in 2015. Since March 2016, he has been working with the School of Information and Communication Engineering, University of Electronic Science and Technology of China (UESTC). He is now a Full Professor and PhD supervisor. He is an editor for \textit{IEEE Wireless Communications Letters}, \textit{IEEE/CIC China Communications} and \textit{Journal of Communications and Information Networks}. He also served for \textit{IEEE Communications Magazine} as a guest editor. He is a program vice-chair for IEEE TrustCom 2020, a technical program committee (TPC) chair for IEEE UCET 2021 and a program vice-chair for UbiSec 2022. He also serves as a TPC member for several prestigious IEEE conferences, such as IEEE Globecom/ICC/WCSP and etc. He has won the outstanding young researcher award of IEEE TCGCC in 2024. He has also won the best paper award of IEEE SustainCom 2020 and the best paper award of IEEE MMTC 2021. His current research focuses on wireless communications and resource management for B5G/6G, wireless information and power transfer as well as integrated communication, computing and sensing.
\end{IEEEbiography}

\end{document}